\documentclass{SciPost}

\hypersetup{
	colorlinks,
	linkcolor={red!50!black},
	citecolor={blue!50!black},
	urlcolor={blue!80!black}
}

\usepackage[bitstream-charter]{mathdesign}
\DeclareSymbolFont{usualmathcal}{OMS}{cmsy}{m}{n}
\DeclareSymbolFontAlphabet{\mathcal}{usualmathcal}

\fancypagestyle{SPstyle}{
	\fancyhf{}
	
	\fancyfoot[C]{\textbf{\thepage}}
}

\usepackage{amsmath}    
\usepackage{graphicx}   
\usepackage{amscd}    
\usepackage{enumitem}
\usepackage{color}
\usepackage{mathrsfs}
\usepackage{bm}
\usepackage{bbm,wasysym}
\usepackage{multirow}
\usepackage{easyReview}
\usepackage{array}
\usepackage{empheq}
\usepackage{pifont}
\usepackage{soul}
\usepackage{svg}
\usepackage{float}

\renewcommand{\todo}[1]{\textcolor{red}{TODO: #1}\ }
\renewcommand{\todo}[1]{}								%%%% make sure to comment this (2/2)

\renewcommand{\(}{\left(}
\renewcommand{\)}{\right)}
\newcommand{\bb}[1]{\mathbb{#1}}

\newcommand{\fc}[1]{\mathcal{#1}}
\newcommand{\scr}[1]{\mathscr{#1}}

\newcommand{\eff}{\text{eff}}

\DeclareMathSymbol{\Alpha}{\mathalpha}{operators}{"41}
\DeclareMathSymbol{\Beta}{\mathalpha}{operators}{"42}
\DeclareMathSymbol{\Epsilon}{\mathalpha}{operators}{"45}
\DeclareMathSymbol{\Zeta}{\mathalpha}{operators}{"5A}
\DeclareMathSymbol{\Eta}{\mathalpha}{operators}{"48}
\DeclareMathSymbol{\Iota}{\mathalpha}{operators}{"49}
\DeclareMathSymbol{\Kappa}{\mathalpha}{operators}{"4B}
\DeclareMathSymbol{\Mu}{\mathalpha}{operators}{"4D}
\DeclareMathSymbol{\Nu}{\mathalpha}{operators}{"4E}
\DeclareMathSymbol{\Omicron}{\mathalpha}{operators}{"4F}
\DeclareMathSymbol{\Rho}{\mathalpha}{operators}{"50}
\DeclareMathSymbol{\Tau}{\mathalpha}{operators}{"54}
\DeclareMathSymbol{\Chi}{\mathalpha}{operators}{"58}
\DeclareMathSymbol{\omicron}{\mathord}{letters}{"6F}

\newcommand{\mc}[1]{\text{\fontencoding{U}\fontfamily{boondoxuprscr}\fontshape{n}\selectfont #1}}

\renewcommand{\d}{\partial}

\newcommand{\8}{\infty}

\newcommand{\p}{\hspace*{5ex}}
\newcommand{\ps}{\hspace*{2ex}}

\newcommand{\<}{\langle}
\renewcommand{\>}{\rangle}

\newcommand{\ec}{\mathtt{e}}

\newcommand{\eps}{\varepsilon}
\newcommand{\cmt}[1]{}

\newcommand{\ee}{\text{E}}

\newcommand{\scri}{\mathscr{I}}

\makeatletter
\newcommand{\oset}[3][0ex]{%
	\mathrel{\mathop{#3}\limits^{
			\vbox to#1{\kern-.5\ex@
				\hbox{$\scriptstyle#2$}\vss}}}}
\makeatother

\makeatletter
\let\old@font@info\@font@info
\def\@font@info#1{%
	\expandafter\ifx\csname\detokenize{#1}\endcsname\relax
	\old@font@info{#1}%
	\fi
	\expandafter\xdef\csname\detokenize{#1}\endcsname{}%
}
\makeatother

\newcommand{\doot}[1]{\overset{\circ}{#1}}

\numberwithin{equation}{subsection}

\begin{document}

\pagestyle{SPstyle}

\begin{center}{\Large \textbf{\color{scipostdeepblue}{
The Energy-Momentum Tensor of a Compact Object in General Relativity to all Multipole Orders
}}}\end{center}

\begin{center}\textbf{
		Trevor A. Scheopner\textsuperscript{1$\star$} and
		Felix A. Lichtner\textsuperscript{1$\dagger$}
}\end{center}

\begin{center}
	{\bf 1} Max Planck Institute for Gravitational Physics (Albert Einstein Institute), Am Mühlenberg 1, Potsdam 14476, Germany
	\\[\baselineskip]
	$\star$ \href{mailto:trevor.scheopner@aei.mpg.de}{\small trevor.scheopner@aei.mpg.de}
	\p
	$\dagger$ \href{mailto:felix.lichtner@aei.mpg.de}{\small felix.lichtner@aei.mpg.de}
\end{center}

\section*{\color{scipostdeepblue}{Abstract}}
\textbf{\boldmath{%
In effective field theory approaches to compact object motion in general relativity, spin and tidal effects are encoded by operators formed from the Riemann tensor and its derivatives. Many such operators contribute degenerately to observables, with a single one often affecting several multipole moments of the body. Using the Mathisson--Papapetrou--Dixon formalism, we construct a natural basis of independent tensors which are each canonically conjugate to precisely one multipole. 
We find the novel exact distributional formula for the energy-momentum tensor of a compact object in terms of its reduced multipole moments and these tensors.
}}

\vspace{\baselineskip}

\vspace{10pt}
\noindent\rule{\textwidth}{1pt}
\tableofcontents
\noindent\rule{\textwidth}{1pt}
\vspace{10pt}

\section{Introduction}
\label{sec:intro}

In the bustling field of gravitational wave physics, ignited by the first observation of gravitational waves by the LIGO/Virgo collaboration~\cite{LIGOScientific:2016aoc, LIGOScientific:2017vwq}, ever more precise theoretical calculations of the motion of compact objects in the general relativistic two-body problem are necessary~\cite{Punturo:2010zz, LISA:2017pwj, Reitze:2019iox, Blanchet:2006zz, Porto:2016pyg, Levi:2018nxp, Buonanno:2022pgc, Kim:2022pou,Kim:2022bwv,Levi:2022dqm,Levi:2022rrq,Mandal:2022nty,Mandal:2022ufb, Goldberger:2009qd}. There are many successful formalisms in active use for the perturbative analytic computation of observables, including in both the post-Newtonian and post-Minkowskian expansions, fully classical methods~\cite{Goldberger:2004jt,Foffa:2012rn,Foffa:2019rdf,Foffa:2016rgu,Foffa:2019yfl,Blumlein:2020pog,Foffa:2019hrb,Blumlein:2021txe,Porto:2024cwd,Blumlein:2021txj,Blumlein:2020znm,Porto:2005ac,Levi:2015msa,Kim:2021rfj,Levi:2020uwu,Levi:2019kgk,Ross:2012fc,Cho:2021mqw,Cho:2022syn,Amalberti:2023ohj, Amalberti:2024jaa,Mandal:2024iug,Dlapa:2021npj,Dlapa:2021vgp,Dlapa:2023hsl,Kalin:2020mvi,Kalin:2020fhe,Kalin:2022hph,Dlapa:2022lmu,Dlapa:2024cje,Dlapa:2025biy,Mougiakakos:2021ckm,Bautista:2021wfy,Bautista:2022wjf, Vines:2017hyw,Siemonsen:2017yux,Porto:2006bt, Faye:2006gx, Damour:2007nc,Liu:2021zxr,Scheopner:2023rzp, Barack:2018yvs, Harte:2009yr, Harte:2008xq, Gralla:2008fg, Poisson:2011nh, Harte:2005yi, Porto:2026fsd, Dlapa:2026oyq, Blanco:2022mgd, Blanco:2023jxf, Blanco:2024fte, Harte:2025tmd, Blanco:2026wwi, Blanco:2026prj, Kim:2025gis, Brunello:2025gpf, Witzany:2026eqc, Skoupy:2026ewu, Albertini:2024agg, Skoupy:2024uan, Witzany:2024ttz, Kerachian:2023oiw, Piovano:2024yks, Witzany:2022vck, Drummond:2023loz, Drummond:2022xej, Drummond:2022efc, deFirmian:2026mln, Compere:2023alp, Ramond:2024sfp, Ramond:2024ozy, Harte:2026ihj}, worldline quantum field theory~\cite{Mogull:2020sak,Jakobsen:2023ndj,Jakobsen:2023pvx,Jakobsen:2022psy,Jakobsen:2023hig,Driesse:2024xad,Driesse:2024feo,Jakobsen:2021smu,Jakobsen:2021zvh, Driesse:2026qiz, Haddad:2025cmw, Mogull:2025cfn}, and several fully field theoretic approaches~\cite{Cheung:2018wkq,Bern:2019nnu,Bern:2022kto,Bern:2024adl,Kosower:2018adc,Bjerrum-Bohr:2018xdl,Cristofoli:2019neg,Damgaard:2019lfh,Brandhuber:2021eyq,Bern:2021dqo,Bern:2021yeh,Damgaard:2023ttc,Bern:2023ccb,Bern:2025zno,Herderschee:2023fxh,Brandhuber:2023hhy,Georgoudis:2023lgf,Cristofoli:2021vyo,Bohnenblust:2023qmy,Bohnenblust:2025gir, Parra-Martinez:2020dzs,Weinzierl:2022eaz,Bern:1994cg,Bern:2011qt, Bern:2010ue,Bern:2019prr,Bern:2022wqg, Chung:2018kqs,Arkani-Hamed:2019ymq,Bern:2020buy, Chiodaroli:2021eug,Aoude:2023vdk,Cangemi:2023ysz,Akpinar:2025bkt,FebresCordero:2022jts, Guevara:2019fsj,Guevara:2018wpp, Kim:2021rda,Kim:2024grz, Cho:2023kux,Damgaard:2024fqj, Bern:2026oqp}. These frameworks all begin with an action which depends on some degrees of freedom for the first body $\Psi_1$, some  for the second $\Psi_2$, and the metric $g_{\alpha\beta}$. The action for the full theory takes the form:
\begin{equation}
	\fc{S}_{\text{full}}[\Psi_1, \Psi_2, g] = \fc{S}_1[\Psi_1, g] + \fc{S}_2[\Psi_2, g] + \fc{S}_{\text{Einstein-Hilbert}}[g]
\end{equation}
where $\fc{S}_{1}$ and $\fc{S}_2$ are the actions for the first and second bodies alone when in placed in an arbitrary metric. The goal of these approaches is to either (perturbatively) directly compute observables (e.g. the radiated gravitational waveform) from this variational principle or to integrate out the gravitational field to produce an action $\fc{S}_{\eff}[\Psi_1,\Psi_2]$ of just the remaining degrees of freedom of the two bodies. In either case, it is the individual actions $\fc{S}_1$ and $\fc{S}_2$ which encode all of the structure of either body and act as the building blocks of the full theory. In this paper we will focus entirely on these individual actions. Thus, we consider a single body placed in a background metric with action $\fc{S}[\Psi, g]$.

We will present the analysis of this paper in the worldline formalism for the sake of concreteness, though our essential conclusions do not make use of special features of the worldline approach and are readily applied to field theoretic methods as well. For a point-particle, $\Psi$ is just the body's worldline $z^\mu(\lambda)$ (parameterized by worldline time $\lambda$) and $\fc{S}$ is $-m$ times the proper time of the body's trajectory. The question of what to write for $\fc{S}[\Psi,g]$ becomes much more interesting as more physics is considered, such as the spin or tidal effects. One takes the standard approach in effective field theory and writes all possible terms that are diffeomorphism scalars in terms of $\Psi$ (which now includes degrees of freedom beyond $z^\mu$) and $g_{\mu\nu}$ (and its derivatives) up to some perturbative order. Generically, this means that the action will include terms containing various powers of the Riemann tensor and its covariant derivatives contracted with body degrees of freedom and evaluated on the worldline, each with arbitrary Wilson coefficients. Scattering amplitudes, from which observables can be extracted, can then be computed as functions of Wilson coefficients to give parameterized precise analytic predictions to a desired order. 

Generically, at any given number of powers of the Riemann tensor and its derivatives, many of the terms one can write down do not contribute independently to physical observables. In typical EFT thinking, such terms are related by use of the lower order equations of motion in the action and so can be reduced by field redefinitions. For example, any term containing the Ricci tensor can be eliminated by field redefinition because (neglecting the effect of the body's self-field), the neighborhood of the body (as it is placed in a background spacetime) is a vacuum solution to Einstein's equations and thus has $R_{\mu\nu} = 0$. As well, many terms can be converted into each other by various integrations by parts. In a worldline action, the combination $\dot z^\mu\nabla_\mu$ acting on a tensor gives a covariant time derivative along the worldline, which can be integrated onto other factors in such a term, mixing Wilson coefficients around at the level of the action. For terms which contain many derivatives and which are beyond linear in the Riemann tensor, it can be difficult to identify by inspection a maximal set of genuinely independent operators. Being able to do so efficiently in terms of a basis of tensors with well-organized transformation properties would be convenient, as having fewer terms in the action with better symmetry-organization results in simpler and more physically clear calculations.

The most fundamental formalism for approaching compact object motion in general relativity is the Mathisson-Papapetrou-Dixon (MPD) theory~\cite{Dixon:1970mpd, dixon2, dixon3}, which provides exact equations of motion for a body's total momentum and spin in terms of its higher mass multipole moments. The theory also provides unique definitions of those higher multipole moments which are \textit{fully reduced}, in the sense of having subtracted out all constraints that result from the covariant conservation of energy-momentum. We introduce the operators canonically conjugate to the reduced field multipole moments and dub them ``multipolar Riemann tensors''. Due to the dynamical independence of the reduced multipoles, the multipolar Riemann tensors form a natural basis of operators to use in the EFT of a compact object which contribute independently to observables.  We find explicit expressions for them and derive a simple, novel distributional formula for the exact energy-momentum tensor of an arbitrary compact object fully in terms of its reduced multipole moments and these multipolar Riemann tensors. 

We begin the paper in Section~\ref{sec:mpd_formalism} with a brief introduction into the physical and mathematical foundations of the MPD formalism, including the assumptions we make about the energy-momentum tensor. We also give an overview of the theory of Synge's worldfunction, bitensors, and of the most relevant definitions and identities surrounding the MPD equations. We then define the multipolar Riemann tensors, show their conjugacy to the general reduced multipoles of the body, and develop a series expression in terms of these quantities for the energy-momentum tensor of a compact object in Section~\ref{sec:mrt}. Next, in Section~\ref{sec:variation}, we show that the MPD equations can be represented by a worldline variational principle to all orders if that action is designed to satisfy a particular orthogonality condition. Under such condition, we show that the momentum, spin, and higher multipole moments of the variational principle are identical to those of the MPD formalism and that the EFT terms within that action are physically independent.  

\section{Overview of the MPD Formalism}
\label{sec:mpd_formalism}

Our analysis depends heavily on the special uniqueness of Dixon's multipole moments. To appreciate their properties, we give a lightning description of the MPD formalism. The authors publish in conjunction with this paper a much more thorough review of this formalism~\cite{mpdreview}.

\subsection{Notation and Basic Conditions}

We always use ``mostly-plus'' signature for the metric and the Levi-Civita connection for the covariant derivative. Consider a completely generic compact body of finite size and arbitrary degrees of freedom in general relativity. The energy-momentum tensor of such a body is necessarily symmetric and covariantly conserved:
\begin{equation}
	T^{\alpha\beta} = T^{\beta\alpha}, \p \nabla_\beta T^{\alpha\beta} = 0  \label{eq:e_cons}
\end{equation}
due to the diffeomorphism invariance of general relativity, regardless of any particular details of the theory by which the body is described\footnote{Black holes are vacuum solutions to Einstein's equations outside of their event horizons, but it is possible to consider effective energy-momentum tensors that give the same solution in the neighborhood of the black hole. Those sources can be made to have support purely within the horizon. See, for example~\cite{Israel:1970kp, Levi:2015msa, Scheopner:2023rzp}. Using these effective sources, black holes can (at least in perturbation theory) be treated by the MPD formalism.}.

Let the worldtube $\bb{W}$ be the convex hull of the support of $T^{\alpha\beta}$. Let the worldline $z^\mu(\lambda)$ be an arbitrary smooth future-orientated timelike curve contained entirely within the worldtube, parameterized by the arbitrary worldline time $\lambda$, which is defined to increase with coordinate time. Let the ``worldline time direction'' $u^\mu(\lambda)$ be an arbitrary differentiable future-orientated timelike unit-vector on the worldline at each worldline time. Let $\Sigma(\lambda)$ be ``space at fixed time'' determined geodesically by an observer on the worldline who assigns $u^\mu$ to be the time direction. That is, $\Sigma(\lambda)$ is the set of points $x^\mu$ in spacetime so that the tangent vector at $z^\mu(\lambda)$ of the geodesic connecting $z^\mu$ to $x^\mu$ is orthogonal to $u^\mu(\lambda)$. The worldline clock $\tau(x)$ is the scalar field giving what time that observer assigns to the event $x$: $\tau(x) = \lambda \iff x \in \Sigma(\lambda)$. Let $\ec(\lambda) = -u\cdot \dot z$ be called the einbein and let $w^\alpha(x)$ be any vector field satisfying $\ec w^\alpha\nabla_\alpha\tau = 1$. $d\Sigma_\alpha$ represents the directed volume measure on $\Sigma$ and $d\Sigma_w = w^\alpha d\Sigma_\alpha$. $\gamma_{\mu\nu} = g_{\mu\nu} + u_\mu u_\nu$ is the induced spatial metric on the worldline. $\doot{f}{}^\Omega = \frac{1}{\ec} \frac{D f^\Omega}{D\lambda}$ is the reparameterization invariant covariant time derivative along the worldline. 

We require the body to be physically reasonable in several ways. (I) For all timelike vector fields $V^\alpha$, $T^{\alpha\beta} V_\beta$ is timelike everywhere that $T^{\alpha\beta} \neq 0$. (II) For all $V^\alpha$ which are timelike, $T^{\alpha\beta} V_\alpha V_\beta \ge 0$ and at least somewhere this is positive. (III) At each worldline time there is precisely one geodesic connecting any two points within $\bb{W}\cap \Sigma(\lambda)$. (IV) There is a bound $\ell$ so that that geodesic distance remains under that bound for all pairs of points in $\bb{W}\cap \Sigma(\lambda)$ and for all time. (VI) The gravitational potential energy of the body in the field it experiences is lesser in magnitude than the body's inertia. Given these conditions, all of the results of the next subsection have been previously proven~\cite{Dixon:1970mpd, dixon2, dixon3, Beiglbock:1965fun, Costa:2014nta} and are reviewed in~\cite{mpdreview}. 

In order to express our results, we need some additional notation. We use beginning Greek alphabet indices ($\alpha, \beta, \gamma, \delta$) for tensors evaluated at a generic spacetime point $x$. We use middle and late Greek alphabet indices ($\mu, \nu, \rho, \sigma, \kappa, \tau, \omega$) for tensors evaluated specifically on the worldline $z$. We will encounter several bitensors, which are functions of two spacetime points ($x$ and $z$) and which can carry indices at each point (each of those indices can only be meaningfully contracted with another index of the same type). We use a capital Greek index $\Omega$ for a tensor with many indices whose particular index structure is unimportant to the equation. That is, $f^\Omega(x) = f^{\alpha_1\dots\alpha_m}{}_{\beta_1\dots\beta_n}$, as a shorthand. The covariant derivative is conveniently written:
\begin{equation}
	\nabla_\alpha f^\Omega = \d_\alpha f^\Omega + \Gamma^\gamma{}_{\alpha\beta} \fc{G}^\beta{}_\gamma f^\Omega
\end{equation}
where $\fc{G}^\beta{}_{\gamma}$ is ``DeWitt's Index Shuffler''~\cite{JanSteinhoff:2015ist} which is a linear operator on tensor fields defined to act as:
\begin{align}
	\fc{G}^\alpha{}_\beta f^{\gamma_1\dots\gamma_m}{}_{\delta_1\dots\delta_n} =& \delta^{\gamma_1}_{\beta} f^{\alpha \gamma_2\dots\gamma_m}{}_{\delta_1\dots\delta_n} + \dots + \delta^{\gamma_m}_{\beta} f^{\gamma_1\dots\gamma_{m-1}\alpha}{}_{\delta_1\dots\delta_n} \nonumber \\
	&\ -\delta^{\alpha}_{\delta_1} f^{\gamma_1\dots\gamma_m}{}_{\beta\delta_2\dots\delta_n} - \dots -\delta^{\alpha}_{\delta_n} f^{\gamma_1\dots\gamma_m}{}_{\delta_1\dots\delta_{n-1}\beta}.
\end{align}
We use angle brackets $\<, \>$ as a shorthand for the inner product of functions:
\begin{equation}
	\<T^{\alpha\beta}, h_{\alpha\beta}\> = \int T^{\alpha\beta}(x) h_{\alpha\beta} D^4 x
\end{equation}
where $D^4 x = \sqrt{-\det g } d^4 x$ always. 

We use $(, )$ to indicate symmetrization of the enclosed indices and $[, ]$ for antisymmetrization, both with a factor $\frac{1}{n!}$ when used on $n$-indices so that they act as the identity transformation when the tensor to which they are applied is already (anti)symmetric. We use $|, |$ to close off indices intended to be excluded from a surrounding (anti)symmetrization and $[', ]'$ to indicate antisymmetrization which is independent of another antisymmetrization ($A^{[\rho['\mu \sigma] \nu]'}$ is antisymmetrized  between $\rho$ and $\sigma$ and independently between $\mu$ and $\nu$). 

$\sigma(x, z)$ is Synge's worldfunction, the biscalar field defined as:
\begin{equation}
	\sigma(x, z) = \frac{1}{2}\int_0^1 g_{\alpha\beta}(\zeta)\frac{d\zeta^\alpha}{ds}\frac{d\zeta^\beta}{ds} ds \label{eq:synge}
\end{equation}
where $\zeta^\alpha(s)$ is always the minimal geodesic connecting $x$ and $z$ in unit affine time: $\zeta^\mu(0) = z^\mu$ and $\zeta^\alpha(1) = x^\alpha$. Synge's worldfunction only exists for $x, z$ so that there is a unique such geodesic. Beginning Greek indices on $\sigma$ indicate covariant $x^\alpha$ derivatives; middle Greek indices indicate covariant $z^\mu$ derivatives. Indices are raised and lowered using the metric at the appropriate point. Define:
\begin{equation}
	v^\mu = -\sigma^\mu(x,z), \p H^\alpha{}_\mu(x,z) = \sigma^{-\alpha}{}_\mu, \p K^\alpha{}_\mu(x,z) = H^\alpha{}_\nu\sigma^\nu{}_\mu
\end{equation}
where $\sigma^{-\alpha}{}_\mu$ is the matrix inverse of $\sigma^\mu{}_\alpha$ (we similarly write $H^{-\mu}{}_\alpha$ to mean the matrix inverse of $H^\alpha{}_\mu$). $v^\mu = -\sigma^\mu$ at fixed $z$ provides a coordinate transformation from the $x^\alpha$ coordinate system to the $v^\mu$ coordinate system, which are Riemann normal coordinates based at $z$. The $x^\alpha$ coordinates are recovered by the exponential map $x = \exp_z(v)$ and $H^\alpha{}_\mu$ serves as the Jacobian matrix of the transformation. The bitensors $H^\alpha{}_\mu$, $K^\alpha{}_\mu$ are the propagators of the geodesic deviation equation in the sense that under a variation in the base point of a geodesic $\delta z^\mu$ and covariant variation its initial tangent vector $\Delta v^\mu$, the corresponding variation in the final point $\delta x^\alpha$ is:
\begin{equation}
	\delta x^\alpha = K^\alpha{}_\mu \delta z^\mu + H^\alpha{}_\mu \Delta v^\mu. 
\end{equation}
The partial derivatives in the $v^\mu$ coordinate system at $z$ define tensor extensions:
\begin{equation}
	\ee_{\rho_1\dots\rho_n} \phi^{\mu\cdots}{}_{\nu\cdots}(z) = \left.\frac{\d^n}{\d v^\rho_1\dots\d v^\rho_n}\(H^{-\mu}{}_\alpha(x(z,v),z)\cdots H^\beta{}_\nu(x(z,v), z)\cdots \phi^{\alpha\cdots}{}_{\beta\cdots}(x(z,v))\)\right|_{v= 0}
\end{equation}
which are the fully symmetric and covariant generalization of the partial derivative to curved spacetime. Tensor extensions are natural in that they are the coefficients of the Taylor series expansion of a tensor expressed in Riemann normal coordinates. Explicit strategies for calculating tensor extensions are given in the review released in conjunction with this paper (see also \cite{Poisson:2011nh}).

\subsection{Results of the MPD Theory}

We now give the results of the above conditions. For a given (at this moment arbitrary) choice of $(z, u)$, the total linear momentum and total spin about the worldline are given by:
\begin{align}
	p_\mu(\lambda) = \int_{\Sigma(\lambda)} T^{\alpha\beta} K_{\beta\mu} d\Sigma_\alpha, \ps \ps S^{\mu\nu}(\lambda) = 2\int_{\Sigma(\lambda)} v^{[\mu} H_\beta{}^{\nu]} T^{\alpha\beta} d\Sigma_\alpha. \label{eq:psdef}
\end{align}
These definitions are unique in that \textit{if} the spacetime and electromagnetic field possess a Killing vector field $\xi^\alpha$, then the combination:
\begin{equation}
	\fc{Q}(\lambda) = p_\mu(\lambda) \xi^\mu(z(\lambda)) + \frac{1}{2} S^{\mu\nu}(\lambda) \nabla_{[\mu}\xi_{\nu]}(z(\lambda)).
\end{equation}
is conserved. In the neighborhood of the worldline $\xi^\mu(z)$ represents the amount of spacetime translation induced by $\xi^\alpha$ while $\nabla_{[\mu}\xi_{\nu]}(z)$ represents the amount of induced Lorentz transformation. Consequently by Noether's theorem, this momentum and spin are precisely the proper generators of these transformations, even in the absence of any symmetries. 

The conservation of energy-momentum \eqref{eq:e_cons} implies, and is implied by, the MPD equations of motion:
\begin{equation}
	\doot p_\mu = - \frac{1}{2}R_{\mu\nu\rho\sigma}\doot z{}^\nu S^{\rho\sigma} + \text{\textbf{F}}_\mu, \p \doot S{}^{\mu\nu} = p^\mu\doot z{}^\nu - \doot z{}^\mu p^\nu + \text{\textbf{T}}^{\mu\nu} \label{eq:mpd}
\end{equation}
where the MPD force $\text{\textbf{F}}_\mu$ and torque $\text{\textbf{T}}^{\mu\nu}$ are given by series expressions in terms of the body's multipole moments. These moments $\fc{J}_n^{\kappa_1\dots\kappa_n \rho\sigma}$, which we call the field mass multipole moments, have time-evolution which is fully unconstrained by \eqref{eq:e_cons} and they fulfill the index symmetries:
\begin{align}
	&\fc{J}_n^{\kappa_1\dots\kappa_n\rho\mu\sigma\nu} = \fc{J}_n^{(\kappa_1\dots\kappa_n)[\rho\mu][\sigma\nu]} = \fc{J}_n^{\kappa_1\dots\kappa_n\sigma\nu\rho\mu}, \p u_{\kappa_1} \fc{J}_n^{\kappa_1\dots\kappa_n\rho\mu\sigma\nu} = 0, \nonumber \\
	&\fc{J}_n^{\kappa_1\dots\kappa_n[\rho\mu\sigma]\nu} = 0, \p \fc{J}_n^{\kappa_1\dots[\kappa_n\rho\mu]\sigma\nu} = 0. \label{eq:indexsym}
\end{align}
These are related to what Dixon calls the reduced multipole moments $M_n^{\kappa_1\cdots\kappa_n\mu\nu}$, and which we call the potential mass multipole moments, via:
\begin{gather}
	M_0^{\mu\nu} = \doot z{}^{(\mu} p^{\nu)}, \p M_1^{\kappa\mu\nu} = S^{\kappa(\mu}\doot z{}^{\nu)}, \p
	M_n^{\kappa_1\dots\kappa_n\mu\nu} = 4\frac{n-1}{n+1} \fc{J}_{n-2}^{(\kappa_1\dots\kappa_{n-1}|\mu|\kappa_n)\nu}. \label{eq:mdef}
\end{gather}
Neglecting $M_0^{\mu\nu}$ and $M_1^{\kappa\mu\nu}$, these two sets of moments are invertibly related. In the gauge theory of linearized gravity, the metric serves as the gauge potential and the linearized Riemann tensor is its gauge invariant field strength. Maintaining this analogy in true general relativity, we refer to the $M_n^{\cdots}$ as \textit{potential} moments because they enter the action naturally through a series over their products with the metric variations of the tensor extensions of the metric, while we refer to the $\fc{J}_n^{\cdots}$ as \textit{field} moments because they enter instead as a series over their products with the metric variations of the multipolar Riemann tensors (as we show in \eqref{eq:gravact}). 

The index symmetries of the field multipole moments ensure they are uniquely given as spatial integrals of the energy-momentum tensor by:
\begin{equation}
	\fc{J}_n^{\kappa_1\dots\kappa_n \rho\mu\sigma\nu} = \mc{t}_{n+2}^{\kappa_1\dots\kappa_n[\rho['\sigma\mu]\nu]'} + \frac{1}{n+1}\mc{p}_{n+2}^{\kappa_1\dots\kappa_n[\rho['\sigma\mu]\nu]'\iota}\doot z{}_\iota \label{eq:curvedj}
\end{equation}
where
\begin{gather}
	\mc{t}_n^{\kappa_1\dots\kappa_n\mu\nu} = \int_\Sigma v^{\kappa_1}\dots v^{\kappa_n} H^{-\mu}{}_\alpha H^{-\nu}{}_\beta T^{\alpha\beta} d\Sigma_w, \ \
	\mc{p}_n^{\kappa_1\dots\kappa_n\mu\nu\sigma} = \int_{\Sigma} 2v^{\kappa_1}\dots v^{\kappa_n} T^{\alpha\beta} H_{\beta\rho} \text{V}_n^{\rho\sigma\mu\nu} d\Sigma_\alpha,	\nonumber 	\\
	\text{V}_n^{\rho\sigma\mu\nu} = (n-1)\int_0^1 \sigma^\rho{}_\alpha(\zeta,z)\sigma^{\alpha(\mu}(\zeta,z) \sigma^{\nu)}{}_\beta(\zeta, z) \sigma^{\beta\sigma}(\zeta,z) s^{n-2} ds. \label{eq:naivedef}
\end{gather}
The quantities $\mc{t}_n$ are the most naive quantities one might take as the multipole moments of the energy-momentum tensor in curved spacetime (they are essentially just powers of displacement from the worldline integrated over a time-slice against $T^{\alpha\beta}$). However, due to the covariant conservation of energy-momentum \eqref{eq:e_cons}, they contain infinitely many complicated interdependencies. The $\fc{J}_n$ quantities are uniquely powerful because beyond the index symmetries \eqref{eq:indexsym}, there are absolutely no constraints on them provided by \eqref{eq:e_cons}. That is, the $\fc{J}_n$ are the genuinely dynamically independent degrees of freedom capturing the higher multipole moments of a generic compact object in general relativity. 

Dixon~\cite{Dixon:1970mpd, dixon2, dixon3} found that, in terms of the potential mass multipole moments, the MPD force and torque are given by:
\begin{align}
	\text{\textbf{F}}_\mu =&\ \frac{1}{2}\sum_{n=2}^\8\frac{1}{n!} M^{\kappa_1\dots \kappa_n \rho\sigma}_n \nabla_\mu \ee_{\kappa_1\dots\kappa_n} g_{\rho\sigma} \nonumber \\
    \text{\textbf{T}}^{\mu\nu} =& \sum_{n=1}^\8 \frac{1}{n!}\(g^{\rho[\mu} M_{n+1}^{\nu] \kappa_1\dots\kappa_n \sigma\lambda} + g^{\sigma[\mu} M_{n+1}^{\nu] \kappa_1\cdots\kappa_n \rho\lambda} - g^{\lambda[\mu} M_{n+1}^{\nu] \kappa_1\cdots\kappa_n\rho\sigma}\) \ee_{\kappa_1\dots\kappa_n\lambda} g_{\rho\sigma}. \label{eq:mpdft}
\end{align}
We show, in \eqref{eq:mpdftj}, how these can be written simply as series in the field mass moments $\fc{J}_n$ and the multipolar Riemann tensors, and observe that the torque can be written much more compactly by using the DeWitt index shuffler. Those simplified series will naturally guide us toward taking a variational principle perspective in \eqref{eq:stressdef}. 

It is possible to choose the time direction $u^\mu$ to coincide with the unit vector in the direction of the momentum and to choose the worldline $z^\mu$ so that the spin represents pure rotations (no boosts) in the center of mass frame, which is stated by the spin supplementary condition (SSC) $S^{\mu\nu} p_\nu = 0$. A detailed treatment of the conditions under which these are possible is given in the accompanying review.

Dixon refers to two equations as the most important in his theory (8.49 and 8.50 in~\cite{dixon3}), which are precisely \eqref{eq:mpd} and the fact that the energy-momentum tensor can be expressed fully in terms of its multipole moments. His expression\footnote{The energy-momentum tensor is written in a slightly different form in his equation 8.49, but using his constructions to evaluate and simplify that equation leads to this form without any noteworthy complication.} gives that for an arbitrary smooth symmetric tensor field of $h_{\alpha\beta}$:
\begin{align}
	\<T^{\alpha\beta}, h_{\alpha\beta}\> =&\ \int_{-\8}^\8\(p^\mu \doot z{}^\nu h_{\mu\nu} + S^{\rho\mu} \doot z{}^\nu \nabla_\rho h_{\mu\nu}\) D\lambda \nonumber \\
	&\ + \int_{-\8}^\8 \sum_{n=2}^\8 \frac{1}{n!} M_n^{\rho_1\dots\rho_n\mu\nu}\left.\frac{\d}{\d v^{\rho_1}\dots \d v^{\rho_n}}\(\bar{h}_{\mu\nu} - L_{\bar\Omega} G_{\mu\nu}\)\right|_{v=0} D\lambda.\label{eq:demt}
\end{align} 
Here $D\lambda = \ec d\lambda$ is the reparameterization invariant worldline time measure, $\bar{h}_{\mu\nu}(z, v)$ is the bitensor defined by:
\begin{equation}
	\bar{h}_{\mu\nu}(z, v) = H^\alpha{}_\mu(x(z, v), z) H^\beta{}_\nu(x(z, v), v) h_{\alpha\beta}(x(z, v))
\end{equation}
(which is just the tensor $h_{\alpha\beta}(x)$ expressed in Riemann normal coordinates $v$ based at $z$), and $G_{\mu\nu}$ is the metric expressed similarly:
\begin{equation}
	G_{\mu\nu}(z, v) = H^\alpha{}_\mu(x(z, v), z) H^\beta{}_\nu(x(z, v), v) g_{\alpha\beta}(x(z, v)). \label{eq:Gdef}
\end{equation}
$L_{\bar\Omega} G_{\mu\nu}$ is the Lie derivative of $G_{\mu\nu}$ with respect to $\bar{\Omega}^\mu$:
\begin{equation}
	L_{\bar\Omega} G_{\mu\nu} = \bar{\Omega}^\rho \frac{\d G_{\mu\nu}}{\d v^\rho} + G_{\mu\rho} \frac{\d \bar{\Omega}^\rho}{\d v^\nu} + G_{\rho\nu} \frac{\d \bar{\Omega}^\rho}{\d v^\mu}.
\end{equation}
$\bar{\Omega}^\mu(z, v)$ is the bitensor $\Omega^\alpha(x, z)$ expressed in Riemann normal coordinates:
\begin{equation}
	\bar{\Omega}^\mu(z, v) = H^{-\mu}{}_\alpha(x(z, v), z) \Omega^\alpha(x(z, v), z)
\end{equation}
and $\Omega^\alpha(x,z)$ is the bitensor which satisfies:
\begin{equation}
	\sigma^\beta(\nabla_\alpha\Omega_\beta + \nabla_\beta \Omega_\alpha) = h_{\alpha\beta}\sigma^\beta, \p \Omega_\alpha(z, z) = 0, \p \delta^\alpha_\mu \delta^\beta_\nu\left.\nabla_\alpha\Omega_\beta\right|_{x\to z} = \frac{1}{2}h_{\mu\nu}(z). \label{eq:omegadef}
\end{equation}
Dixon proves the existence and uniqueness of $\Omega^\alpha$. Dixon's \eqref{eq:demt} is difficult to use to infer the behavior of the energy-momentum tensor as a distribution because of $\Omega$'s complicated functional dependence on $h_{\alpha\beta}$. One of our main results will be to simplify \eqref{eq:demt} into a novel and much more insightful form, giving an exact expression in full general relativity of the behavior of the energy-momentum distribution as a series in its reduced multipoles contracted into derivatives (tensor extensions) of the metric on the body's effective worldline.\footnote{Equation 6.36 of~\cite{Scheopner:2023rzp} attempts to accomplish this but ultimately expresses the energy-momentum tensor in terms of the non-reduced quantities given in its equation 6.33. We will improve on this formula.}. 

\section{Multipolar Riemann Tensors}
\label{sec:mrt}

\subsection{Definition and Properties}

We now introduce a set of tensors that will greatly simplify equation \eqref{eq:demt} and provide much clearer contact with variational principle based approaches to compact object motion (which were not considered in Dixon's original analysis). The simplest way to motivate the definition of these multipolar Riemann tensors is to start with equation \eqref{eq:mpdft} for the force and torque, and to rewrite them using the field mass moments. Doing so via \eqref{eq:mdef} (using the DeWitt index shuffler) leads to:
\begin{align}
	\text{\textbf{F}}_\mu =&\  - \sum_{n=0}^\8\frac{n+1}{(n+3)!} \fc{J}^{\kappa_1\dots\kappa_n\rho\tau\sigma\omega}_n \nabla_\mu R^{(n)}_{\kappa_1\dots\kappa_n\rho\tau\sigma\omega}  \nonumber \\
	\text{\textbf{T}}^{\mu\nu} =&\ 2\sum_{n=0}^\8\frac{n+1}{(n+3)!} \fc{J}^{\kappa_1\dots\kappa_n\rho\tau\sigma\omega}_n\fc{G}^{[\mu\nu]} R^{(n)}_{\kappa_1\dots\kappa_n\rho\tau\sigma\omega} \label{eq:mpdftj}
\end{align}
where the multipolar Riemann tensors $R^{(n)}_{\cdots}$ are defined by:
\begin{equation}
	R^{(n)}_{\kappa_1\dots\kappa_n\rho\mu\sigma\nu}(z) = -2\ee_{\kappa_1\dots\kappa_n [\rho['\sigma}g_{\mu]\nu]'}.  \label{eq:mrtdef}
\end{equation}

The first few multipolar Riemann tensors are given explicitly by:
\begin{align}
	& \p \p R^{(0)}_{\rho\mu\sigma\nu} = R_{\rho\mu\sigma\nu}, \p R^{(1)}_{\kappa\rho\mu\sigma\nu} = \nabla_\kappa R_{\rho\mu\sigma\nu}, \nonumber \\
	R^{(2)}_{\kappa\lambda\rho\mu\sigma\nu} = &\ \nabla^2_{(\kappa\lambda)} R_{\rho\mu\sigma\nu} + \frac{8}{9} R^\tau{}_{(\rho\sigma)(\kappa} R_{\lambda)(\mu\nu) \tau} - \frac{8}{9} R^\tau{}_{(\mu\sigma)(\kappa}R_{\lambda)(\rho\nu)\tau} \nonumber \\
	& + \frac{1}{3}R^\tau{}_{\mu\sigma\nu} R_{\tau(\kappa\lambda)\rho}  - \frac{1}{3}R^\tau{}_{\rho\sigma\nu} R_{\tau(\kappa\lambda)\mu} + \frac{1}{3} R^\tau{}_{\nu\rho\mu} R_{\tau(\kappa\lambda)\sigma} - \frac{1}{3} R^\tau{}_{\sigma\rho\mu} R_{\tau(\kappa\lambda)\nu}. 
\end{align}
In general, $R^{(n)}_{\kappa_1\dots\kappa_n\rho\mu\sigma\nu} = \nabla^n_{(\kappa_1\dots\kappa_n)} R_{\rho\mu\sigma\nu} + $ corrections with at least two factors of the Riemann tensor and at most $n-2$ total covariant derivatives on those factors. These additional terms serve an essential purpose. The symmetrized covariant derivatives of the Riemann tensor do not satisfy a differential Bianchi identity in the sense of the last of \eqref{eq:indexsym}, while the multipolar Riemann tensors do. Thus, the multipolar Riemann tensors have all of the symmetries necessary to be conjugate to the field mass moments in curved spacetime. In particular:
\begin{equation}
	R^{(n)}_{\kappa_1\dots\kappa_n\rho\mu\sigma\nu} = R^{(n)}_{(\kappa_1\dots\kappa_n)[\rho\mu][\sigma\nu]} = R^{(n)}_{\kappa_1\dots\kappa_n\sigma\nu\rho\mu}, \ps R^{(n)}_{\kappa_1\dots\kappa_n[\rho\mu\sigma]\nu} = 0, \ps R^{(n)}_{\kappa_1\dots[\kappa_n\rho\mu]\sigma\nu} = 0.  \label{eq:mrtsyms}
\end{equation}

\subsection{Exact Energy-Momentum Tensor to all Multipole Orders}

The key insight that will allow us to improve on \eqref{eq:demt} is to reframe the distributional behavior of the energy-momentum tensor through the use of a variational principle. In particular, in curved spacetime, if our body's dynamics are described by an action $\fc{S}[\Psi, g]$, then its energy-momentum tensor is given by the metric variation of that action under a compactly supported perturbation as: 
\begin{equation}
	\delta \fc{S} = \frac{1}{2}\int T^{\alpha\beta} \delta g_{\alpha\beta} D^4 x = \frac{1}{2}\<T^{\alpha\beta}, \delta g_{\alpha\beta}\>. \label{eq:stressdef}
\end{equation}
In consideration of this, it becomes natural in \eqref{eq:demt} to think of $h_{\alpha\beta} = \delta g_{\alpha\beta}$ as the perturbation of the metric $g_{\alpha\beta}$ and to use this thinking to simplify $\bar h_{\mu\nu}$ and $L_{\bar\Omega} G_{\mu\nu}$ appearing on the right hand side of that equation. Equation \eqref{eq:demt}, neglecting the contribution from $L_{\bar{\Omega}} G_{\mu\nu}$, becomes a series then in the tensor extensions of the metric variation $\ee_{\rho_1\dots \rho_n} h_{\mu\nu}$. We will show that when the $L_{\bar{\Omega}} G_{\mu\nu}$ contributions are included, the result instead becomes a series in the metric variations of the tensor extensions of the full metric, i.e. $\delta \ee_{\rho_1\dots\rho_n} g_{\mu\nu}$. 

We begin by calculating $\Omega_\alpha$. Consider a metric perturbation applied to Synge's worldfunction. Directly varying \eqref{eq:synge} with respect to the metric at fixed end points, we find:
\begin{equation}
	\delta\sigma(x,z) = \frac{1}{2}\int_0^1 h_{\alpha\beta}(\zeta)\dot\zeta^\alpha\dot\zeta^\beta ds.  \label{eq:deltasig}
\end{equation}
Note that the geodesic path $\zeta^\alpha(s)$ between $z$ and $x$ is itself varied, so that the varied path satisfies the geodesic equation with the varied metric, but due to $\sigma$ being on-shell with respect to the geodesic equation its net contribution vanishes. Contracting \eqref{eq:omegadef} with $\sigma^\alpha$ and using our standard geodesic $\zeta^\alpha(s)$, we find:
\begin{equation}
	\frac{d}{ds}(\dot\zeta^\alpha\Omega_\alpha) = \frac{1}{2}h_{\alpha\beta}\dot\zeta^\alpha\dot\zeta^\beta \implies \sigma^\alpha(x,z)\Omega_\alpha(x,z) =  \delta\sigma(x,z) \label{eq:omegadeltasig}
\end{equation}
using the first boundary condition in \eqref{eq:omegadef}. The $z$ derivative of this gives:    
\begin{equation}
	\nabla_\mu\delta\sigma = \sigma^\alpha{}_\mu\Omega_\alpha + \sigma^\alpha\nabla_\mu\Omega_\alpha. \label{eq:omegasigder}
\end{equation}
We can compute $\sigma^\alpha\nabla_\mu\Omega_\alpha$ by taking the $z$ derivative of \eqref{eq:omegadef} and contracting the result with $\sigma^\alpha$. Doing so gives (again using our standard geodesic):
\begin{equation}
	\sigma^\alpha\sigma^\beta\nabla_\alpha \nabla_\mu \Omega_\beta = 0 \implies \frac{D}{Ds}(\dot\zeta^\alpha \nabla_\mu \Omega_\alpha) = 0 \implies \sigma^\alpha\nabla_\mu\Omega_\alpha = v^\nu \delta^\alpha_\nu\left. \nabla_\mu\Omega_\alpha\right|_{x\to z}.  \label{eq:omegabc}
\end{equation}
We can determine the $z$ derivative of $\Omega_\alpha$ at coincidence by taking the $z$ derivative of the first boundary condition in \eqref{eq:omegadef} and then using the second boundary condition to simplify:
\begin{equation}
	\delta^\beta_\nu\left.\nabla_\beta\Omega_\alpha\right|_{x\to z} + \left.\nabla_\nu\Omega_\alpha\right|_{x\to z} = 0 \implies \delta^\alpha_\nu \left.\nabla_\mu\Omega_\alpha\right|_{x\to z} = -\frac{1}{2}h_{\mu\nu}(z). \label{eq:omegader}
\end{equation}
Returning \eqref{eq:omegader} to \eqref{eq:omegabc}, then taking that into \eqref{eq:omegasigder} gives:
\begin{equation}
	\bar{\Omega}^\mu = H^{-\mu}{}_\alpha\Omega^\alpha = -\nabla^\mu\delta\sigma - \frac{1}{2} h^{\mu\nu}(z) v_\nu. \label{eq:omegabarsol}
\end{equation}
The right hand side here is now an explicit functional of $h_{\alpha\beta}$ through \eqref{eq:deltasig} which we will use to simplify \eqref{eq:demt}. 

Now let us study the metric variation of $G_{\mu\nu}$ as defined in \eqref{eq:Gdef}. By varying $G_{\mu\nu}(z,v)$ at fixed $(z, v)$, we find by using the chain rule:
\begin{equation}
	\delta G_{\mu\nu} = \delta H^\alpha{}_\mu H^\beta{}_\nu g_{\alpha\beta} + H^\alpha{}_\mu \delta H^\beta{}_\nu g_{\alpha\beta} + H^\alpha{}_\mu H^\beta{}_\nu \delta g_{\alpha\beta} + \delta x^\gamma\nabla_\gamma (H^\alpha{}_\mu H^\beta{}_\nu g_{\alpha\beta}) \label{eq:deltabigg}
\end{equation}
where the variations in $H^\alpha{}_\mu$ and $g_{\alpha\beta}$ are taken at fixed $(x,z)$ and $\delta x^\gamma$ is the variation in $x^\gamma$ under a change in the metric at fixed $(z,v)$. From $H^\alpha{}_\mu = -\sigma^{-\alpha}{}_\mu$, it immediately follows that:
\begin{equation}
	\delta H^\alpha{}_\mu = H^\alpha{}_\nu \bar{\d}_\mu \delta\sigma^\nu
\end{equation}
where $\delta\sigma^\nu$ is the metric variation of $\sigma^\nu$. Because the metric is varied, $\delta \sigma^\mu \neq \nabla^\mu \delta\sigma$ as $\sigma^\mu = g^{\mu\nu}\nabla_\nu \sigma$. Instead:
\begin{equation}
	\delta\sigma^\mu = \nabla^\mu\delta\sigma + h^{\mu\nu}v_\nu \label{eq:deltasignab}
\end{equation}
after using $v^\mu = -\sigma^\mu(x,z)$. By differentiating $v^\mu = -\sigma^\mu(x,z)$ it immediately follows that at fixed $(z,v)$:
\begin{equation}
	\delta x^\alpha = H^\alpha{}_\mu \delta \sigma^\mu. 
\end{equation}
Returning these to \eqref{eq:deltabigg}, we find:
\begin{equation}
	\delta G_{\mu\nu}(z,v) = \bar{h}_{\mu\nu} + \delta \sigma^\rho \bar{\d}_\rho G_{\mu\nu} + G_{\mu\rho}\bar{\d}_\nu\delta\sigma^\rho + G_{\rho\nu}\bar{\d}_\mu\delta\sigma^\rho. 
\end{equation}
Using \eqref{eq:deltasignab} and \eqref{eq:omegabarsol}, this becomes:
\begin{equation}
	\bar{h}_{\mu\nu} - L_{\bar{\Omega}} G_{\mu\nu} = \delta G_{\mu\nu}(z,v) - \frac{1}{2}h^{\rho\sigma} v_\rho \bar{\d}_\sigma G_{\mu\nu} - \frac{1}{2}h^\rho{}_\mu G_{\rho\nu} - \frac{1}{2}h^\rho{}_\nu G_{\rho\mu}.  \label{eq:diximp}
\end{equation}
We are now ready to feed this into \eqref{eq:demt}. To do so, note that the metric variation of the extensions of the metric satisfy, by definition:
\begin{equation}
	\delta \ee_{\rho_1\dots\rho_n} g_{\mu\nu} = \left.\frac{\d^n \delta G_{\mu\nu}}{\d v^{\rho_1}\dots v^{\rho_n}}\right|_{v= 0}. 
\end{equation}
Then, \eqref{eq:diximp} put into \eqref{eq:demt} and rewriting using the multipolar Riemann tensors gives:
\begin{align}
	\delta \fc{S} =&\ \frac{1}{2}\int T^{\alpha\beta} \delta g_{\alpha\beta} D^4 x \nonumber \\
	=&\ \frac{1}{2}\int_{-\8}^\8\(p^\mu \doot z{}^\nu \delta g_{\mu\nu} + S^{\rho\mu} \doot z{}^\nu \nabla_\rho \delta g_{\mu\nu}\) D\lambda \nonumber \\
	&\ - \int_{-\8}^\8 \sum_{n=0}^\8 \frac{n+1}{(n+3)!} \fc{J}_n^{\kappa_1\dots\kappa_n\rho\mu\sigma\nu}\(\delta R^{(n)}_{\kappa_1\dots\kappa_n\rho\mu\sigma\nu} + \frac{1}{2} \delta g_{\tau\omega}\fc{G}^{\tau\omega} R^{(n)}_{\kappa_1\dots\kappa_n\rho\mu\sigma\nu}\) D\lambda. \label{eq:gravact}
\end{align}
This is the proper full simplification of \eqref{eq:demt}. Equation \eqref{eq:gravact} fully determines $T^{\alpha\beta}$ as a distribution, as it specifies how it behaves when integrated against arbitrary test functions. If the energy-momentum tensor of a compact object is given a distributional expression on a worldline, then (due to the uniqueness of Dixon's moments satisfying \eqref{eq:indexsym}), equation \eqref{eq:gravact} allows one to easily read off the momentum, spin, and multipole moments of the compact object, just by looking at each term in the series of that distributional expression. Equation \eqref{eq:gravact} generalizes equation 2.13 of~\cite{Saketh:2022wap} by being valid at all multipole orders and improves upon equation 6.36 of~\cite{Scheopner:2023rzp} by being written in terms of the body's reduced multipole moments. By explicitly expanding the tensor extensions to the desired order, \eqref{eq:gravact} can be used to extract the energy-momentum tensor as a series over covariant derivatives of the Dirac delta as in equation 2.13 of~\cite{Saketh:2022wap}, though the use of a Dirac delta distribution in curved spacetime is highly formal.  

\section{Worldline Variational Principles}
\label{sec:variation}

Equation \eqref{eq:gravact} is ideal for making contact to a worldline action. The use of a worldline variational principle for the motion of compact objects is well-established~\cite{JanSteinhoff:2015ist,Levi:2015msa,Levi:2018nxp,Vines:2017hyw}. The most general worldline action, allowing arbitrary internal structure, can be written in the form $\fc{S}[z, p, \Lambda, S, \scri, \scr{P}, \ec, \beta, \chi]$. Here $z^\mu$ is the worldline and $p_\mu$ is the body's momentum. $\Lambda^\mu{}_I$ is a tetrad tracking the frame of the body (so that $\Lambda^\mu{}_I \Lambda^\nu{}_J \eta^{IJ} = g^{\mu\nu}$ under all variations and with angular velocity $\Omega^{\mu\nu} = \eta^{IJ} \Lambda^\mu{}_I \frac{D\Lambda^\nu{}_J}{D\lambda}$ with capital Latin indices being Lorentz indices in the body-frame). $S^{\mu\nu}$ is the spin tensor of the body (satisfying the covariant SSC $S^{\mu\nu} p_\nu = 0$). $\scri^\Omega$ are internal degrees of freedom of the body with conjugate momenta $\scr{P}_\Omega$. $\ec$ is the einbein (which is a Lagrange multiplier enforcing reparameterization invariance), $\beta_\mu$ is a Lagrange multiplier enforcing the SSC, and $\chi_\mu$ is a Lagrange multiplier enforcing that the body's time direction is defined as the unit vector in the direction of its momentum ($\Lambda^\mu{}_0 = u^\mu$). Such an action then takes the form~\cite{JanSteinhoff:2015ist}:
\begin{equation}
	\fc{S} = \int_{-\8}^\8\(p_\mu\dot z^\mu + \frac{1}{2}S^{\mu\nu}\Omega_{\mu\nu} + \scr{P}_\Omega\dot\scri^\Omega + \ec\(\sqrt{|p^2|} - \fc{M}\) + \beta_{\mu} S^{\mu\nu} p_\nu + \chi_{\mu}(\Lambda^\mu{}_0 - u^\mu)\)d\lambda \label{eq:wlact}
\end{equation}
where the dynamical mass function $\fc{M}(g, \Phi, u, \Lambda, S, \scri, \scr{P})$ is a scalar function (that is invariant under a change of coordinates) that depends on $z$ only through the metric on the worldline $g_{\mu\nu}(z)$ and whatever other tensors $\Phi$ on the worldline the body is interacting with (such as the Riemann tensor and its derivatives). We define the internal body degrees of freedom $\scr{I}^\Omega$ and $\scr{P}_\Omega$ to be in the body frame so that they are invariant under coordinate diffeomorphisms (i.e. all indices are internal Lorentz indices, which can be accomplished through use of the body tetrad). 

In order to compute the energy-momentum tensor due to \eqref{eq:wlact}, we will have to vary $\fc{S}$ with respect to the metric. So, it matters whether we consider the dynamical mass function to depend on $u^\mu$ or on $u_\mu$ (and similarly so for the index placements on other tensors). We, by convention, take it as a function of $u^\mu, \Lambda^\mu{}_i$, $S^{\mu\nu}$ with all spacetime indices raised. Since $\fc{M}$ must be invariant under a change of coordinates, we can consider the coordinate transformation $x'^\alpha = x^\alpha + \eps\xi^\alpha(x)$ and enforce that the $\eps$ derivative of $\fc{M}$ at $\eps = 0$ must vanish. Doing so, because this must hold for any $\xi^\alpha$, we find:
\begin{equation}
	2\frac{\d\fc{M}}{\d g_{\mu\nu}} = \frac{\d\fc{M}}{\d \Phi^\Omega} \fc{G}^{\mu\nu} \Phi^\Omega +  u^\mu \frac{\d \fc{M}}{\d u_\nu} + \Lambda^\mu{}_i \frac{\d\fc{M}}{\d \Lambda_{\nu i}} + 2S^\mu{}_\rho\frac{\d\fc{M}}{\d S_{\nu\rho}}. \label{eq:scalarity}
\end{equation}
The are no terms due to $\scri, \scr{P}$ because we have by definition expressed these degrees of freedom in the body frame and so they carry no spacetime indices. Using \eqref{eq:scalarity} we can always eliminate the metric derivative of the dynamical mass function from any other equation. Additionally, the left hand side is automatically symmetric, so the antisymmetrization of the right hand side vanishes. 

We suggestively define the ``provisional force'' $\text{\textbf{F}}_\mu$ and ``provisional torque'' $\text{\textbf{T}}^{\mu\nu}$ by:
\begin{equation}
	\text{\textbf{F}}_\mu = -\frac{\d\fc{M}}{\d\Phi^\Omega} \nabla_\mu \Phi^\Omega \p \text{\textbf{T}}^{\mu\nu} = 2\frac{\d\fc{M}}{\d\Phi^\Omega} \fc{G}^{[\mu\nu]} \Phi^\Omega.  \label{eq:provt}
\end{equation}
We will show shortly that these coincide with the definitions in \eqref{eq:mpdftj}. By \eqref{eq:scalarity}:
\begin{equation}
    \text{\textbf{T}}^{\mu\nu} = \frac{\d\fc{M}}{\d u_\mu} u^\nu -  \frac{\d\fc{M}}{\d u_\nu} u^\mu +  \frac{\d\fc{M}}{\d \Lambda_{\mu i}} \Lambda^\nu{}_i -  \frac{\d\fc{M}}{\d \Lambda_{\nu i} } \Lambda^\mu{}_i + 2\frac{\d\fc{M}}{\d S_{\mu\rho}}S^\nu{}_\rho - 2\frac{\d\fc{M}}{\d S_{\nu\rho}} S^\mu{}_\rho. 
\end{equation}
Varying the action \eqref{eq:wlact} with respect to all of its dependencies except the metric, requiring the result to be stationary, using those conditions to eliminate all Lagrange multipliers, and simplifying the results yields the equations of motion:
\begin{align}
	&\fc{M}\doot z{}^\mu = p^\mu - \text{\textbf{T}}{}^{\mu\nu} u_\nu - S^{\mu\nu} \doot u_\nu, &
	&\doot p{}_\mu = -\frac{1}{2} R_{\mu\nu\rho\sigma} \doot z{}^\nu S^{\rho\sigma} + q\fc{F}_{\mu\nu}\doot z{}^\nu + \text{\textbf{F}}_\mu, &\nonumber \\
	&\doot \Lambda{}^\mu{}_i = \(u^\mu \doot u{}^\nu + 2\gamma^{\mu\rho}\gamma^{\nu\sigma}\frac{\d\fc{M}}{\d S^{\rho\sigma}}\) \Lambda_{\nu i}, &
	&\doot S{}^{\mu\nu} = p^\mu \doot z{}^\nu - p^\nu \doot z{}^\mu + \text{\textbf{T}}{}^{\mu\nu},&\nonumber \\
	&\doot{\scri}{}^\Omega = \frac{\d\fc{M}}{\d \scr{P}_{\Omega}},&
	&\doot{\scr{P}}{}_\Omega = -\frac{\d\fc{M}}{\d \scri^\Omega}. \label{eq:shorteom}
\end{align}
The $p$ and $S$ equations of motion are almost identical to the MPD equations \eqref{eq:mpd}! They would be the same if we had justified identifying the provisional force and torque with the MPD force and torque. This will be accomplished through the energy-momentum tensor. The equations of motion in \eqref{eq:shorteom} beyond those mirroring the MPD equations tell us how the center of mass worldline is determined, the body frame rotates, and the internal degrees of freedom evolve. 

Now let us consider the metric variation of \eqref{eq:gravact} so that we can extract the energy-momentum tensor it implies. Doing so and using the equations of motion \eqref{eq:shorteom} to simplify, we find:
\begin{equation}
	\delta \fc{S} = \frac{1}{2}\int_{-\8}^\8\(\ p^\mu\doot z{}^\nu \delta g_{\mu\nu} + S^{\mu\nu} \doot z{}^\rho \nabla_\mu \delta g_{\nu\rho}  -2 \sum_{n=0}^\8 \frac{\d\fc{M}}{\d \Phi^\Omega}\bigl(\delta \Phi^\Omega + \frac{1}{2} \delta g_{\mu\nu}\fc{G}^{\mu\nu} \Phi^\Omega\bigr) \) D\lambda  \label{eq:variact}
\end{equation}
where $\delta \Phi^\Omega$ is the metric variation of $\Phi^\Omega$. Equation \eqref{eq:variact} holds regardless of what fields $\Phi^\Omega$ exist external to the worldline. Because we are in particular considering the interaction of the worldline with gravity, we take $\Phi^\Omega$ to be all of the tensors that can be formed out of the metric as a function evaluated at $z$. By default one might take the Riemann tensor and all orders of its covariant derivatives, however, at each order in differentiation we can exchange the covariant derivatives of the Riemann tensor with the multipolar Riemann tensors up to that same order. Thus, we just as well take $\Phi^\Omega$ to be the list of multipolar Riemann tensors at all orders. Doing so, and owing precisely to the uniqueness of Dixon's multipole moments in satisfying their index symmetries, we can use this choice when comparing \eqref{eq:gravact} to identify that the worldline momentum and spin are exactly Dixon's momentum and spin and that:
\begin{equation}
	\fc{J}_n^{\kappa_1\dots\kappa_n \rho\mu\sigma\nu} = (n+2)(n+3) n! \frac{\d\fc{M}}{\d R^{(n)}_{\kappa_1\dots\kappa_n\rho\mu\sigma\nu}} \label{eq:jfromdmf}
\end{equation}
where, due to \eqref{eq:indexsym}, the dynamical mass function must satisfy the orthogonality relation:
\begin{equation}
	u_{\kappa_1} \frac{\d\fc{M}}{\d R^{(n)}_{\kappa_1\dots\kappa_n\rho\mu\sigma\nu}} = 0. \label{eq:dmfortho}
\end{equation}
This orthogonality relation is not a condition on physics; it is a condition on how the dynamical mass function must be written in order for the coefficients in \eqref{eq:variact} to correspond to fully reduced multipole moments. 

Dynamical mass functions which differ in how they depend on derivatives in the time direction of the Riemann tensor may produce the same physics, because those time derivatives can be iteratively eliminated through use of the equations of motion at each order. Such different expressions will give different integrands at each $\lambda$ in \eqref{eq:variact} even though the integrated result is unchanged. Because any covariantly conserved energy-momentum tensor can be put into the form of \eqref{eq:gravact} \textit{uniquely}, we can use that form to uniquely fix how time-derivatives of the Riemann tensor appear in the dynamical mass function, the result being \eqref{eq:dmfortho}. With these identifications, our worldline equations of motion reproduce the entirety of the MPD formalism. 

The direct and exact identification between quantities in the worldline formalism and in the MPD formalism has profound implications for effective field theory approaches to the motion of compact objects. Ordinarily in effective field theory one has enormous perturbative field redefinition invariance available for consideration, which allows the significant mixing of terms at the level of the action. Dixon proved that the $\fc{J}_n$ moments exist and are unique given their index symmetries \eqref{eq:indexsym}. Therefore, it is always possible to enforce \eqref{eq:dmfortho} when constructing the dynamical mass function. The resulting $z$, $p$, and $S$ are uniquely defined. In other words, \eqref{eq:dmfortho} can be viewed as a field gauge choice for the worldline degrees of freedom that performs a complete such gauge fixing. Additionally that gauge fixing has a clear physical interpretation in terms of the true multipole moments of the body. In particular, the resulting momentum and spin are the true Noether charges generating translations and Lorentz transformations in the neighborhood of the worldline and the higher multipole moments are fully dynamically unconstrained. 

\section{Conclusion}

We achieved a simple formula, valid to all multipole orders, for the energy-momentum tensor of a compact object in terms of its fully reduced field multipole moments $\fc{J}_n$ in \eqref{eq:gravact}. The simplifications from the MPD theory to this formula were made possible by taking the perspective of treating the test function against which the energy-momentum tensor is integrated when thought of as a distribution as being a perturbation of the metric of the spacetime in which the body moves. The result is far simpler than that in the traditional MPD theory \eqref{eq:demt} and much easier to use when matching to an effective field theory description for the compact object. 

There are many subtleties to formulating EFTs for spinning compact objects in curved spacetime. Direct enumerations of EFT operators often produce terms which contribute degenerately to amplitudes and observables, due to them being able to be related using lower order equations of motion and field redefinition. The reduced multipole moments of the MPD formalism provide a direct pathway to developing a maximally efficiently organized variational principle in which the terms have clear physical interpretations due to their uniqueness.

While an EFT can be written using the covariant derivatives of the Riemann tensor directly, these covariant derivatives fail to manifest the differential Bianchi identity past the first order. This is reflected in the worldline quantities to which they are conjugate being not fully dynamically reduced, i.e. they are still partly constrained by the covariant conservation of energy-momentum. The multipolar Riemann tensors introduced in \eqref{eq:mrtdef} manifest all of the index symmetries of the Riemann tensor and generalize the differential Bianchi identity to all orders, as seen in \eqref{eq:mrtsyms}. As a consequence, they are conjugate precisely to the fully reduced $\fc{J}_n$ field multipole moments of the body. In order for a worldline EFT to manifestly match these reduced moments, the dynamical mass function of the theory must be designed to satisfy \eqref{eq:dmfortho}. This has the effect of providing an a priori mechanism to unique select your field definitions gauge and thus ensure that every remaining operator written down contributes non-degenerately to some physical observable. 

The multipolar Riemann tensors provide a reduction advantage over the covariant derivatives of the Riemann tensor, but they may still be decomposed further. The next natural step would be to take the decomposition of the Riemann tensor into the Weyl tensor (in vacuum) and then into the electric and magnetic tidal tensors, and to generalize it to all orders in the multipolar Riemann tensors. We leave this as a direction for future work.

The fundamental question of what is the proper dynamical mass function for the effective worldline of a spinning black hole in general relativity remains unsolved at high orders in the Riemann tensor. At linear order in the Riemann tensor (and its derivatives), the answer is provided to all orders in spin by matching to the Kerr solution~\cite{Levi:2015msa, Vines:2017hyw, Arkani-Hamed:2017jhn}. At quadratic order in the Riemann tensor (and its derivatives), the answer is in principle provided by matching to the results of the Teukolsky equation, though doing so at high orders in spin is difficult~\cite{Bjerrum-Bohr:2026low, Bautista:2021wfy, Bautista:2022wjf, Bautista:2023sdf, Bautista:2024agp}. Alternatively, many quadratic in Riemann proposals to all orders in spin have been given in literature based on various symmetry principles~\cite{Aoude:2022trd, Bern:2022kto, Cangemi:2023bpe, Scheopner:2023rzp, Bjerrum-Bohr:2023iey, Guevara:2018wpp, Akpinar:2025tct, Akpinar:2025bkt, Akpinar:2024meg, Bern:2020buy, Aoude:2025xxq, Aoude:2023vdk, Guevara:2020xjx, Ben-Shahar:2025tiz, Alessio:2025nzd, Bohnenblust:2024hkw, Cangemi:2022bew, Chiodaroli:2021eug, Cangemi:2023ysz, FebresCordero:2022jts, Guevara:2019fsj}. Because black holes are expected to be the simplest objects in nature~\cite{1983mtbh.book.....C}, we expect that the effective worldline action for a spinning black hole ought to be exceptionally simple as well, when expressed in the proper basis. Because the multipole moments of a Kerr black hole take such a simple form~\cite{Scheopner:2023rzp, vanVoorthuizen:2024goj}, we intend to explore in future work how their conjugates, the multipolar Riemann tensors, may be used to provide that basis. 

\section*{Acknowledgements}
The authors would like to thank Francisco Blanco, Richard Myers, Raj Patil, M.V.S. Saketh, Jan Steinhoff, and Justin Vines for valuable discussions regarding this work. No LLMs or other generative AI tools were used in the science or writing of this work. 

\paragraph{Funding information}

T.S. is supported by ERC grant GWSky / 101167314, funded by the European Union. Views and opinions expressed are however those of the authors only and do not necessarily reflect those of the European Union or the European Research Council Executive Agency. Neither the European Union nor the granting authority can be held responsible for them.


\begin{thebibliography}{100}
	\providecommand{\url}[1]{\texttt{#1}}
	\providecommand{\urlprefix}{URL }
	\expandafter\ifx\csname urlstyle\endcsname\relax
	\providecommand{\doi}[1]{doi:\discretionary{}{}{}#1}\else
	\providecommand{\doi}{doi:\discretionary{}{}{}\begingroup \urlstyle{rm}\Url}\fi
	\providecommand{\eprint}[2][]{\url{#2}}
	
	\bibitem{LIGOScientific:2016aoc}
	B.~P. Abbott \emph{et~al.},
	\newblock \emph{{Observation of Gravitational Waves from a Binary Black Hole Merger}},
	\newblock Phys. Rev. Lett. \textbf{116}(6), 061102 (2016),
	\newblock \doi{10.1103/PhysRevLett.116.061102},
	\newblock \eprint{1602.03837}.
	
	\bibitem{LIGOScientific:2017vwq}
	B.~P. Abbott \emph{et~al.},
	\newblock \emph{{GW170817: Observation of Gravitational Waves from a Binary Neutron Star Inspiral}},
	\newblock Phys. Rev. Lett. \textbf{119}(16), 161101 (2017),
	\newblock \doi{10.1103/PhysRevLett.119.161101},
	\newblock \eprint{1710.05832}.
	
	\bibitem{Punturo:2010zz}
	M.~Punturo \emph{et~al.},
	\newblock \emph{{The Einstein Telescope: A third-generation gravitational wave observatory}},
	\newblock Class. Quant. Grav. \textbf{27}, 194002 (2010),
	\newblock \doi{10.1088/0264-9381/27/19/194002}.
	
	\bibitem{LISA:2017pwj}
	P.~Amaro-Seoane \emph{et~al.},
	\newblock \emph{{Laser Interferometer Space Antenna}}  (2017),
	\newblock \eprint{1702.00786}.
	
	\bibitem{Reitze:2019iox}
	D.~Reitze \emph{et~al.},
	\newblock \emph{{Cosmic Explorer: The U.S. Contribution to Gravitational-Wave Astronomy beyond LIGO}},
	\newblock Bull. Am. Astron. Soc. \textbf{51}(7), 035 (2019),
	\newblock \eprint{1907.04833}.
	
	\bibitem{Blanchet:2006zz}
	L.~Blanchet,
	\newblock \emph{{Gravitational radiation from post-Newtonian sources and inspiralling compact binaries}},
	\newblock Living Rev. Rel. \textbf{9}, 4 (2006).
	
	\bibitem{Porto:2016pyg}
	R.~A. Porto,
	\newblock \emph{{The effective field theorist{\textquoteright}s approach to gravitational dynamics}},
	\newblock Phys. Rept. \textbf{633}, 1 (2016),
	\newblock \doi{10.1016/j.physrep.2016.04.003},
	\newblock \eprint{1601.04914}.
	
	\bibitem{Levi:2018nxp}
	M.~Levi,
	\newblock \emph{{Effective Field Theories of Post-Newtonian Gravity: A comprehensive review}},
	\newblock Rept. Prog. Phys. \textbf{83}(7), 075901 (2020),
	\newblock \doi{10.1088/1361-6633/ab12bc},
	\newblock \eprint{1807.01699}.
	
	\bibitem{Buonanno:2022pgc}
	A.~Buonanno, M.~Khalil, D.~O'Connell, R.~Roiban, M.~P. Solon and M.~Zeng,
	\newblock \emph{{Snowmass White Paper: Gravitational Waves and Scattering Amplitudes}},
	\newblock In \emph{{Snowmass 2021}} (2022), \eprint{2204.05194}.
	
	\bibitem{Kim:2022pou}
	J.-W. Kim, M.~Levi and Z.~Yin,
	\newblock \emph{{N$^{3}$LO spin-orbit interaction via the EFT of spinning gravitating objects}},
	\newblock JHEP \textbf{05}, 184 (2023),
	\newblock \doi{10.1007/JHEP05(2023)184},
	\newblock \eprint{2208.14949}.
	
	\bibitem{Kim:2022bwv}
	J.-W. Kim, M.~Levi and Z.~Yin,
	\newblock \emph{{N$^{3}$LO quadratic-in-spin interactions for generic compact binaries}},
	\newblock JHEP \textbf{03}, 098 (2023),
	\newblock \doi{10.1007/JHEP03(2023)098},
	\newblock \eprint{2209.09235}.
	
	\bibitem{Levi:2022dqm}
	M.~Levi, R.~Morales and Z.~Yin,
	\newblock \emph{{From the EFT of spinning gravitating objects to Poincar{\'e} and gauge invariance at the 4.5PN precision frontier}},
	\newblock JHEP \textbf{09}, 090 (2023),
	\newblock \doi{10.1007/JHEP09(2023)090},
	\newblock \eprint{2210.17538}.
	
	\bibitem{Levi:2022rrq}
	M.~Levi and Z.~Yin,
	\newblock \emph{{Completing the fifth PN precision frontier via the EFT of spinning gravitating objects}},
	\newblock JHEP \textbf{04}, 079 (2023),
	\newblock \doi{10.1007/JHEP04(2023)079},
	\newblock \eprint{2211.14018}.
	
	\bibitem{Mandal:2022nty}
	M.~K. Mandal, P.~Mastrolia, R.~Patil and J.~Steinhoff,
	\newblock \emph{{Gravitational spin-orbit Hamiltonian at NNNLO in the post-Newtonian framework}},
	\newblock JHEP \textbf{03}, 130 (2023),
	\newblock \doi{10.1007/JHEP03(2023)130},
	\newblock \eprint{2209.00611}.
	
	\bibitem{Mandal:2022ufb}
	M.~K. Mandal, P.~Mastrolia, R.~Patil and J.~Steinhoff,
	\newblock \emph{{Gravitational quadratic-in-spin Hamiltonian at NNNLO in the post-Newtonian framework}},
	\newblock JHEP \textbf{07}, 128 (2023),
	\newblock \doi{10.1007/JHEP07(2023)128},
	\newblock \eprint{2210.09176}.
	
	\bibitem{Goldberger:2009qd}
	W.~D. Goldberger and A.~Ross,
	\newblock \emph{{Gravitational radiative corrections from effective field theory}},
	\newblock Phys. Rev. D \textbf{81}, 124015 (2010),
	\newblock \doi{10.1103/PhysRevD.81.124015},
	\newblock \eprint{0912.4254}.
	
	\bibitem{Goldberger:2004jt}
	W.~D. Goldberger and I.~Z. Rothstein,
	\newblock \emph{{An Effective field theory of gravity for extended objects}},
	\newblock Phys. Rev. D \textbf{73}, 104029 (2006),
	\newblock \doi{10.1103/PhysRevD.73.104029},
	\newblock \eprint{hep-th/0409156}.
	
	\bibitem{Foffa:2012rn}
	S.~Foffa and R.~Sturani,
	\newblock \emph{{Dynamics of the gravitational two-body problem at fourth post-Newtonian order and at quadratic order in the Newton constant}},
	\newblock Phys. Rev. D \textbf{87}(6), 064011 (2013),
	\newblock \doi{10.1103/PhysRevD.87.064011},
	\newblock \eprint{1206.7087}.
	
	\bibitem{Foffa:2019rdf}
	S.~Foffa and R.~Sturani,
	\newblock \emph{{Conservative dynamics of binary systems to fourth Post-Newtonian order in the EFT approach I: Regularized Lagrangian}},
	\newblock Phys. Rev. D \textbf{100}(2), 024047 (2019),
	\newblock \doi{10.1103/PhysRevD.100.024047},
	\newblock \eprint{1903.05113}.
	
	\bibitem{Foffa:2016rgu}
	S.~Foffa, P.~Mastrolia, R.~Sturani and C.~Sturm,
	\newblock \emph{{Effective field theory approach to the gravitational two-body dynamics, at fourth post-Newtonian order and quintic in the Newton constant}},
	\newblock Phys. Rev. D \textbf{95}(10), 104009 (2017),
	\newblock \doi{10.1103/PhysRevD.95.104009},
	\newblock \eprint{1612.00482}.
	
	\bibitem{Foffa:2019yfl}
	S.~Foffa, R.~A. Porto, I.~Rothstein and R.~Sturani,
	\newblock \emph{{Conservative dynamics of binary systems to fourth Post-Newtonian order in the EFT approach II: Renormalized Lagrangian}},
	\newblock Phys. Rev. D \textbf{100}(2), 024048 (2019),
	\newblock \doi{10.1103/PhysRevD.100.024048},
	\newblock \eprint{1903.05118}.
	
	\bibitem{Blumlein:2020pog}
	J.~Bl{\"u}mlein, A.~Maier, P.~Marquard and G.~Sch{\"a}fer,
	\newblock \emph{{Fourth post-Newtonian Hamiltonian dynamics of two-body systems from an effective field theory approach}},
	\newblock Nucl. Phys. B \textbf{955}, 115041 (2020),
	\newblock \doi{10.1016/j.nuclphysb.2020.115041},
	\newblock \eprint{2003.01692}.
	
	\bibitem{Foffa:2019hrb}
	S.~Foffa, P.~Mastrolia, R.~Sturani, C.~Sturm and W.~J. Torres~Bobadilla,
	\newblock \emph{{Static two-body potential at fifth post-Newtonian order}},
	\newblock Phys. Rev. Lett. \textbf{122}(24), 241605 (2019),
	\newblock \doi{10.1103/PhysRevLett.122.241605},
	\newblock \eprint{1902.10571}.
	
	\bibitem{Blumlein:2021txe}
	J.~Bl{\"u}mlein, A.~Maier, P.~Marquard and G.~Sch{\"a}fer,
	\newblock \emph{{The fifth-order post-Newtonian Hamiltonian dynamics of two-body systems from an effective field theory approach}},
	\newblock Nucl. Phys. B \textbf{983}, 115900 (2022),
	\newblock \doi{10.1016/j.nuclphysb.2022.115900},
	\newblock [Erratum: Nucl.Phys.B 985, 115991 (2022)],
	\newblock \eprint{2110.13822}.
	
	\bibitem{Porto:2024cwd}
	R.~A. Porto, M.~M. Riva and Z.~Yang,
	\newblock \emph{{Nonlinear gravitational radiation reaction: failed tail, memories {\&} squares}},
	\newblock JHEP \textbf{04}, 050 (2025),
	\newblock \doi{10.1007/JHEP04(2025)050},
	\newblock \eprint{2409.05860}.
	
	\bibitem{Blumlein:2021txj}
	J.~Bl\"umlein, A.~Maier, P.~Marquard and G.~Sch\"afer,
	\newblock \emph{{The 6th post-Newtonian potential terms at $O(G_N^4)$}},
	\newblock Phys. Lett. B \textbf{816}, 136260 (2021),
	\newblock \doi{10.1016/j.physletb.2021.136260},
	\newblock \eprint{2101.08630}.
	
	\bibitem{Blumlein:2020znm}
	J.~Bl{\"u}mlein, A.~Maier, P.~Marquard and G.~Sch{\"a}fer,
	\newblock \emph{{Testing binary dynamics in gravity at the sixth post-Newtonian level}},
	\newblock Phys. Lett. B \textbf{807}, 135496 (2020),
	\newblock \doi{10.1016/j.physletb.2020.135496},
	\newblock \eprint{2003.07145}.
	
	\bibitem{Porto:2005ac}
	R.~A. Porto,
	\newblock \emph{{Post-Newtonian corrections to the motion of spinning bodies in NRGR}},
	\newblock Phys. Rev. D \textbf{73}, 104031 (2006),
	\newblock \doi{10.1103/PhysRevD.73.104031},
	\newblock \eprint{gr-qc/0511061}.
	
	\bibitem{Levi:2015msa}
	M.~Levi and J.~Steinhoff,
	\newblock \emph{{Spinning gravitating objects in the effective field theory in the post-Newtonian scheme}},
	\newblock JHEP \textbf{09}, 219 (2015),
	\newblock \doi{10.1007/JHEP09(2015)219},
	\newblock \eprint{1501.04956}.
	
	\bibitem{Kim:2021rfj}
	J.-W. Kim, M.~Levi and Z.~Yin,
	\newblock \emph{{Quadratic-in-spin interactions at fifth post-Newtonian order probe new physics}},
	\newblock Phys. Lett. B \textbf{834}, 137410 (2022),
	\newblock \doi{10.1016/j.physletb.2022.137410},
	\newblock \eprint{2112.01509}.
	
	\bibitem{Levi:2020uwu}
	M.~Levi, A.~J. Mcleod and M.~Von~Hippel,
	\newblock \emph{{N$^{3}$LO gravitational quadratic-in-spin interactions at G$^{4}$}},
	\newblock JHEP \textbf{07}, 116 (2021),
	\newblock \doi{10.1007/JHEP07(2021)116},
	\newblock \eprint{2003.07890}.
	
	\bibitem{Levi:2019kgk}
	M.~Levi, S.~Mougiakakos and M.~Vieira,
	\newblock \emph{{Gravitational cubic-in-spin interaction at the next-to-leading post-Newtonian order}},
	\newblock JHEP \textbf{01}, 036 (2021),
	\newblock \doi{10.1007/JHEP01(2021)036},
	\newblock \eprint{1912.06276}.
	
	\bibitem{Ross:2012fc}
	A.~Ross,
	\newblock \emph{{Multipole expansion at the level of the action}},
	\newblock Phys. Rev. D \textbf{85}, 125033 (2012),
	\newblock \doi{10.1103/PhysRevD.85.125033},
	\newblock \eprint{1202.4750}.
	
	\bibitem{Cho:2021mqw}
	G.~Cho, B.~Pardo and R.~A. Porto,
	\newblock \emph{{Gravitational radiation from inspiralling compact objects: Spin-spin effects completed at the next-to-leading post-Newtonian order}},
	\newblock Phys. Rev. D \textbf{104}(2), 024037 (2021),
	\newblock \doi{10.1103/PhysRevD.104.024037},
	\newblock \eprint{2103.14612}.
	
	\bibitem{Cho:2022syn}
	G.~Cho, R.~A. Porto and Z.~Yang,
	\newblock \emph{{Gravitational radiation from inspiralling compact objects: Spin effects to the fourth post-Newtonian order}},
	\newblock Phys. Rev. D \textbf{106}(10), L101501 (2022),
	\newblock \doi{10.1103/PhysRevD.106.L101501},
	\newblock \eprint{2201.05138}.
	
	\bibitem{Amalberti:2023ohj}
	L.~Amalberti, F.~Larrouturou and Z.~Yang,
	\newblock \emph{{Multipole expansion at the level of the action in d-dimensions}},
	\newblock Phys. Rev. D \textbf{109}(10), 104027 (2024),
	\newblock \doi{10.1103/PhysRevD.109.104027},
	\newblock \eprint{2312.02868}.
	
	\bibitem{Amalberti:2024jaa}
	L.~Amalberti, Z.~Yang and R.~A. Porto,
	\newblock \emph{{Gravitational radiation from inspiralling compact binaries to N3LO in the effective field theory approach}},
	\newblock Phys. Rev. D \textbf{110}(4), 044046 (2024),
	\newblock \doi{10.1103/PhysRevD.110.044046},
	\newblock \eprint{2406.03457}.
	
	\bibitem{Mandal:2024iug}
	M.~K. Mandal, P.~Mastrolia, R.~Patil and J.~Steinhoff,
	\newblock \emph{{Radiating Love: adiabatic tidal fluxes and modes up to next-to-next-to-leading post-Newtonian order}},
	\newblock JHEP \textbf{05}, 008 (2025),
	\newblock \doi{10.1007/JHEP05(2025)008},
	\newblock \eprint{2412.01706}.
	
	\bibitem{Dlapa:2021npj}
	C.~Dlapa, G.~K{\"a}lin, Z.~Liu and R.~A. Porto,
	\newblock \emph{{Dynamics of binary systems to fourth Post-Minkowskian order from the effective field theory approach}},
	\newblock Phys. Lett. B \textbf{831}, 137203 (2022),
	\newblock \doi{10.1016/j.physletb.2022.137203},
	\newblock \eprint{2106.08276}.
	
	\bibitem{Dlapa:2021vgp}
	C.~Dlapa, G.~K\"alin, Z.~Liu and R.~A. Porto,
	\newblock \emph{{Conservative Dynamics of Binary Systems at Fourth Post-Minkowskian Order in the Large-Eccentricity Expansion}},
	\newblock Phys. Rev. Lett. \textbf{128}(16), 161104 (2022),
	\newblock \doi{10.1103/PhysRevLett.128.161104},
	\newblock \eprint{2112.11296}.
	
	\bibitem{Dlapa:2023hsl}
	C.~Dlapa, G.~K{\"a}lin, Z.~Liu and R.~A. Porto,
	\newblock \emph{{Bootstrapping the relativistic two-body problem}},
	\newblock JHEP \textbf{08}, 109 (2023),
	\newblock \doi{10.1007/JHEP08(2023)109},
	\newblock \eprint{2304.01275}.
	
	\bibitem{Kalin:2020mvi}
	G.~K\"alin and R.~A. Porto,
	\newblock \emph{{Post-Minkowskian Effective Field Theory for Conservative Binary Dynamics}},
	\newblock JHEP \textbf{11}, 106 (2020),
	\newblock \doi{10.1007/JHEP11(2020)106},
	\newblock \eprint{2006.01184}.
	
	\bibitem{Kalin:2020fhe}
	G.~K\"alin, Z.~Liu and R.~A. Porto,
	\newblock \emph{{Conservative Dynamics of Binary Systems to Third Post-Minkowskian Order from the Effective Field Theory Approach}},
	\newblock Phys. Rev. Lett. \textbf{125}(26), 261103 (2020),
	\newblock \doi{10.1103/PhysRevLett.125.261103},
	\newblock \eprint{2007.04977}.
	
	\bibitem{Kalin:2022hph}
	G.~K{\"a}lin, J.~Neef and R.~A. Porto,
	\newblock \emph{{Radiation-reaction in the Effective Field Theory approach to Post-Minkowskian dynamics}},
	\newblock JHEP \textbf{01}, 140 (2023),
	\newblock \doi{10.1007/JHEP01(2023)140},
	\newblock \eprint{2207.00580}.
	
	\bibitem{Dlapa:2022lmu}
	C.~Dlapa, G.~K{\"a}lin, Z.~Liu, J.~Neef and R.~A. Porto,
	\newblock \emph{{Radiation Reaction and Gravitational Waves at Fourth Post-Minkowskian Order}},
	\newblock Phys. Rev. Lett. \textbf{130}(10), 101401 (2023),
	\newblock \doi{10.1103/PhysRevLett.130.101401},
	\newblock \eprint{2210.05541}.
	
	\bibitem{Dlapa:2024cje}
	C.~Dlapa, G.~K{\"a}lin, Z.~Liu and R.~A. Porto,
	\newblock \emph{{Local in Time Conservative Binary Dynamics at Fourth Post-Minkowskian Order}},
	\newblock Phys. Rev. Lett. \textbf{132}(22), 221401 (2024),
	\newblock \doi{10.1103/PhysRevLett.132.221401},
	\newblock \eprint{2403.04853}.
	
	\bibitem{Dlapa:2025biy}
	C.~Dlapa, G.~K{\"a}lin, Z.~Liu and R.~A. Porto,
	\newblock \emph{{Local-in-Time Conservative Binary Dynamics at Fifth Post-Minkowskian and First Self-Force Orders}}  (2025),
	\newblock \eprint{2506.20665}.
	
	\bibitem{Mougiakakos:2021ckm}
	S.~Mougiakakos, M.~M. Riva and F.~Vernizzi,
	\newblock \emph{{Gravitational Bremsstrahlung in the post-Minkowskian effective field theory}},
	\newblock Phys. Rev. D \textbf{104}(2), 024041 (2021),
	\newblock \doi{10.1103/PhysRevD.104.024041},
	\newblock \eprint{2102.08339}.
	
	\bibitem{Bautista:2021wfy}
	Y.~F. Bautista, A.~Guevara, C.~Kavanagh and J.~Vines,
	\newblock \emph{{Scattering in black hole backgrounds and higher-spin amplitudes. Part I}},
	\newblock JHEP \textbf{03}, 136 (2023),
	\newblock \doi{10.1007/JHEP03(2023)136},
	\newblock \eprint{2107.10179}.
	
	\bibitem{Bautista:2022wjf}
	Y.~F. Bautista, A.~Guevara, C.~Kavanagh and J.~Vines,
	\newblock \emph{{Scattering in black hole backgrounds and higher-spin amplitudes. Part II}},
	\newblock JHEP \textbf{05}, 211 (2023),
	\newblock \doi{10.1007/JHEP05(2023)211},
	\newblock \eprint{2212.07965}.
	
	\bibitem{Vines:2017hyw}
	J.~Vines,
	\newblock \emph{{Scattering of two spinning black holes in post-Minkowskian gravity, to all orders in spin, and effective-one-body mappings}},
	\newblock Class. Quant. Grav. \textbf{35}(8), 084002 (2018),
	\newblock \doi{10.1088/1361-6382/aaa3a8},
	\newblock \eprint{1709.06016}.
	
	\bibitem{Siemonsen:2017yux}
	N.~Siemonsen, J.~Steinhoff and J.~Vines,
	\newblock \emph{{Gravitational waves from spinning binary black holes at the leading post-Newtonian orders at all orders in spin}},
	\newblock Phys. Rev. D \textbf{97}(12), 124046 (2018),
	\newblock \doi{10.1103/PhysRevD.97.124046},
	\newblock \eprint{1712.08603}.
	
	\bibitem{Porto:2006bt}
	R.~A. Porto and I.~Z. Rothstein,
	\newblock \emph{{The Hyperfine Einstein-Infeld-Hoffmann potential}},
	\newblock Phys. Rev. Lett. \textbf{97}, 021101 (2006),
	\newblock \doi{10.1103/PhysRevLett.97.021101},
	\newblock \eprint{gr-qc/0604099}.
	
	\bibitem{Faye:2006gx}
	G.~Faye, L.~Blanchet and A.~Buonanno,
	\newblock \emph{{Higher-order spin effects in the dynamics of compact binaries. I. Equations of motion}},
	\newblock Phys. Rev. D \textbf{74}, 104033 (2006),
	\newblock \doi{10.1103/PhysRevD.74.104033},
	\newblock \eprint{gr-qc/0605139}.
	
	\bibitem{Damour:2007nc}
	T.~Damour, P.~Jaranowski and G.~Schaefer,
	\newblock \emph{{Hamiltonian of two spinning compact bodies with next-to-leading order gravitational spin-orbit coupling}},
	\newblock Phys. Rev. D \textbf{77}, 064032 (2008),
	\newblock \doi{10.1103/PhysRevD.77.064032},
	\newblock \eprint{0711.1048}.
	
	\bibitem{Liu:2021zxr}
	Z.~Liu, R.~A. Porto and Z.~Yang,
	\newblock \emph{{Spin Effects in the Effective Field Theory Approach to Post-Minkowskian Conservative Dynamics}},
	\newblock JHEP \textbf{06}, 012 (2021),
	\newblock \doi{10.1007/JHEP06(2021)012},
	\newblock \eprint{2102.10059}.
	
	\bibitem{Scheopner:2023rzp}
	T.~Scheopner and J.~Vines,
	\newblock \emph{{Dynamical Implications of the Kerr Multipole Moments for Spinning Black Holes}}  (2023),
	\newblock \eprint{2311.18421}.
	
	\bibitem{Barack:2018yvs}
	L.~Barack and A.~Pound,
	\newblock \emph{{Self-force and radiation reaction in general relativity}},
	\newblock Rept. Prog. Phys. \textbf{82}(1), 016904 (2019),
	\newblock \doi{10.1088/1361-6633/aae552},
	\newblock \eprint{1805.10385}.
	
	\bibitem{Harte:2009yr}
	A.~I. Harte,
	\newblock \emph{{Effective stress-energy tensors, self-force, and broken symmetry}},
	\newblock Class. Quant. Grav. \textbf{27}, 135002 (2010),
	\newblock \doi{10.1088/0264-9381/27/13/135002},
	\newblock \eprint{0910.4614}.
	
	\bibitem{Harte:2008xq}
	A.~I. Harte,
	\newblock \emph{{Self-forces from generalized Killing fields}},
	\newblock Class. Quant. Grav. \textbf{25}, 235020 (2008),
	\newblock \doi{10.1088/0264-9381/25/23/235020},
	\newblock \eprint{0807.1150}.
	
	\bibitem{Gralla:2008fg}
	S.~E. Gralla and R.~M. Wald,
	\newblock \emph{{A Rigorous Derivation of Gravitational Self-force}},
	\newblock Class. Quant. Grav. \textbf{25}, 205009 (2008),
	\newblock \doi{10.1088/0264-9381/25/20/205009},
	\newblock [Erratum: Class.Quant.Grav. 28, 159501 (2011)],
	\newblock \eprint{0806.3293}.
	
	\bibitem{Poisson:2011nh}
	E.~Poisson, A.~Pound and I.~Vega,
	\newblock \emph{{The Motion of point particles in curved spacetime}},
	\newblock Living Rev. Rel. \textbf{14}, 7 (2011),
	\newblock \doi{10.12942/lrr-2011-7},
	\newblock \eprint{1102.0529}.
	
	\bibitem{Harte:2005yi}
	A.~Harte,
	\newblock \emph{{Self-forces on extended bodies in electrodynamics}},
	\newblock Phys. Rev. D \textbf{73}, 065006 (2006),
	\newblock \doi{10.1103/PhysRevD.73.065006},
	\newblock \eprint{gr-qc/0508123}.
	
	\bibitem{Porto:2026fsd}
	R.~A. Porto and M.~M. Riva,
	\newblock \emph{{Black hole dynamics at fifth post-Newtonian order}},
	\newblock JHEP \textbf{09}, 075 (2026),
	\newblock \doi{10.1007/JHEP09(2026)075},
	\newblock \eprint{2604.09545}.
	
	\bibitem{Dlapa:2026oyq}
	C.~Dlapa, G.~K{\"a}lin, Z.~Liu and R.~A. Porto,
	\newblock \emph{{Nonlocal-in-time tail effects in gravitational scattering to fifth post-Minkowskian and tenth self-force orders}},
	\newblock Phys. Rev. D \textbf{114}(2), 024029 (2026),
	\newblock \doi{10.1103/wkp4-vy6g},
	\newblock \eprint{2604.25916}.
	
	\bibitem{Blanco:2022mgd}
	F.~M. Blanco and {\'E}.~{\'E}. Flanagan,
	\newblock \emph{{Particle Motion under the Conservative Piece of the Self-Force is Hamiltonian}},
	\newblock Phys. Rev. Lett. \textbf{130}(5), 051201 (2023),
	\newblock \doi{10.1103/PhysRevLett.130.051201},
	\newblock \eprint{2205.01667}.
	
	\bibitem{Blanco:2023jxf}
	F.~M. Blanco and {\'E}.~{\'E}. Flanagan,
	\newblock \emph{{Motion of a spinning particle under the conservative piece of the self-force is Hamiltonian to first order in mass and spin}},
	\newblock Phys. Rev. D \textbf{107}(12), 124017 (2023),
	\newblock \doi{10.1103/PhysRevD.107.124017},
	\newblock \eprint{2302.10233}.
	
	\bibitem{Blanco:2024fte}
	F.~M. Blanco,
	\newblock \emph{{Local Hamiltonian dynamics from nonlocal action principles and applications to binary systems in general relativity}},
	\newblock Phys. Rev. D \textbf{110}(2), 024061 (2024),
	\newblock \doi{10.1103/PhysRevD.110.024061},
	\newblock \eprint{2406.00106}.
	
	\bibitem{Harte:2025tmd}
	A.~I. Harte, F.~M. Blanco and E.~E. Flanagan,
	\newblock \emph{{Nonlinearly Self-Interacting Extended Bodies Move as Test Bodies in Effective External Fields}},
	\newblock Phys. Rev. Lett. \textbf{135}(15), 151401 (2025),
	\newblock \doi{10.1103/k74d-dj7y},
	\newblock \eprint{2504.11912}.
	
	\bibitem{Blanco:2026wwi}
	F.~M. Blanco, E.~E. Flanagan and A.~I. Harte,
	\newblock \emph{{Conservative and dissipative sectors in a nonlinear scalar model for the gravitational self-force problem}}  (2026),
	\newblock \eprint{2605.14958}.
	
	\bibitem{Blanco:2026prj}
	F.~M. Blanco,
	\newblock \emph{{The Magnusian generator for dissipative systems and application to leading 2.5PN radiation-reaction dynamics}},
	\newblock JHEP \textbf{10}, 210 (2026),
	\newblock \doi{10.1007/JHEP10(2026)210},
	\newblock \eprint{2607.24335}.
	
	\bibitem{Kim:2025gis}
	J.-W. Kim, R.~Patil, T.~Scheopner and J.~Steinhoff,
	\newblock \emph{{Magnusian: relating the eikonal phase, the on-shell action, and the scattering generator}},
	\newblock JHEP \textbf{03}, 241 (2026),
	\newblock \doi{10.1007/JHEP03(2026)241},
	\newblock \eprint{2511.05649}.
	
	\bibitem{Brunello:2025gpf}
	G.~Brunello, M.~K. Mandal, P.~Mastrolia, R.~Patil, M.~Pegorin, J.~Ronca, S.~Smith, J.~Steinhoff and W.~J. Torres~Bobadilla,
	\newblock \emph{{Six-Loop Gravitational Interactions at the Sixth Post-Newtonian Order}},
	\newblock Phys. Rev. Lett. \textbf{137}(11), 111401 (2026),
	\newblock \doi{10.1103/c2r3-cqtf},
	\newblock \eprint{2512.19498}.
	
	\bibitem{Witzany:2026eqc}
	V.~Witzany and V.~Skoup{\'y},
	\newblock \emph{{Separability of the motion of spinning test particles in curved space-time}}  (2026),
	\newblock \eprint{2606.25792}.
	
	\bibitem{Skoupy:2026ewu}
	V.~Skoup{\'y},
	\newblock \emph{{New approach to the calculation of extreme-mass-ratio inspirals with a spinning secondary}},
	\newblock Phys. Rev. D \textbf{114}(8), 084030 (2026),
	\newblock \doi{10.1103/kpj2-bftm},
	\newblock \eprint{2603.13482}.
	
	\bibitem{Albertini:2024agg}
	A.~Albertini, V.~Skoup{\'y}, G.~Lukes-Gerakopoulos and A.~Nagar,
	\newblock \emph{{Comparing effective-one-body and Mathisson-Papapetrou-Dixon results for a spinning test particle on circular equatorial orbits around a Kerr black hole}},
	\newblock Phys. Rev. D \textbf{111}(6), 064086 (2025),
	\newblock \doi{10.1103/PhysRevD.111.064086},
	\newblock \eprint{2412.16077}.
	
	\bibitem{Skoupy:2024uan}
	V.~Skoup{\'y} and V.~Witzany,
	\newblock \emph{{Analytic Solution for the Motion of Spinning Particles in Kerr Spacetime}},
	\newblock Phys. Rev. Lett. \textbf{134}(17), 171401 (2025),
	\newblock \doi{10.1103/PhysRevLett.134.171401},
	\newblock \eprint{2411.16855}.
	
	\bibitem{Witzany:2024ttz}
	V.~Witzany, V.~Skoup{\'y}, L.~C. Stein and S.~Tanay,
	\newblock \emph{{Actions of spinning compact binaries: Spinning particle in Kerr matched to dynamics at 1.5 post-Newtonian order}},
	\newblock Phys. Rev. D \textbf{111}(4), 044032 (2025),
	\newblock \doi{10.1103/PhysRevD.111.044032},
	\newblock \eprint{2411.09742}.
	
	\bibitem{Kerachian:2023oiw}
	M.~Kerachian, L.~Polcar, V.~Skoup{\'y}, C.~Efthymiopoulos and G.~Lukes-Gerakopoulos,
	\newblock \emph{{Action-angle formalism for extreme mass ratio inspirals in Kerr spacetime}},
	\newblock Phys. Rev. D \textbf{108}(4), 044004 (2023),
	\newblock \doi{10.1103/PhysRevD.108.044004},
	\newblock \eprint{2301.08150}.
	
	\bibitem{Piovano:2024yks}
	G.~A. Piovano, C.~Pantelidou, J.~Mac~Uilliam and V.~Witzany,
	\newblock \emph{{Spinning particles near Kerr black holes: Orbits and gravitational-wave fluxes through the Hamilton-Jacobi formalism}},
	\newblock Phys. Rev. D \textbf{111}(4), 044009 (2025),
	\newblock \doi{10.1103/PhysRevD.111.044009},
	\newblock \eprint{2410.05769}.
	
	\bibitem{Witzany:2022vck}
	V.~Witzany,
	\newblock \emph{{Action-angle coordinates for black-hole geodesics I: Spherically symmetric and Schwarzschild}}  (2022),
	\newblock \eprint{2203.11952}.
	
	\bibitem{Drummond:2023loz}
	L.~V. Drummond, A.~G. Hanselman, D.~R. Becker and S.~A. Hughes,
	\newblock \emph{{Extreme mass-ratio inspiral of a spinning body into a Kerr black hole I: Evolution along generic trajectories}}  (2023),
	\newblock \eprint{2305.08919}.
	
	\bibitem{Drummond:2022xej}
	L.~V. Drummond and S.~A. Hughes,
	\newblock \emph{{Precisely computing bound orbits of spinning bodies around black holes. I. General framework and results for nearly equatorial orbits}},
	\newblock Phys. Rev. D \textbf{105}(12), 124040 (2022),
	\newblock \doi{10.1103/PhysRevD.105.124040},
	\newblock \eprint{2201.13334}.
	
	\bibitem{Drummond:2022efc}
	L.~V. Drummond and S.~A. Hughes,
	\newblock \emph{{Precisely computing bound orbits of spinning bodies around black holes. II. Generic orbits}},
	\newblock Phys. Rev. D \textbf{105}(12), 124041 (2022),
	\newblock \doi{10.1103/PhysRevD.105.124041},
	\newblock \eprint{2201.13335}.
	
	\bibitem{deFirmian:2026mln}
	C.~de~Firmian and J.~Vines,
	\newblock \emph{{Generalized Carter {\&} R{\"u}diger Constants of $\sqrt{\text{Kerr}}$}}  (2026),
	\newblock \eprint{2602.18790}.
	
	\bibitem{Compere:2023alp}
	G.~Comp{\`e}re, A.~Druart and J.~Vines,
	\newblock \emph{{Generalized Carter constant for quadrupolar test bodies in Kerr spacetime}},
	\newblock SciPost Phys. \textbf{15}(6), 226 (2023),
	\newblock \doi{10.21468/SciPostPhys.15.6.226},
	\newblock \eprint{2302.14549}.
	
	\bibitem{Ramond:2024sfp}
	P.~Ramond and S.~Isoyama,
	\newblock \emph{{Symplectic mechanics of relativistic spinning compact bodies: Canonical formalism in the Schwarzschild spacetime}},
	\newblock Phys. Rev. D \textbf{111}(6), 064027 (2025),
	\newblock \doi{10.1103/PhysRevD.111.064027},
	\newblock \eprint{2402.05049}.
	
	\bibitem{Ramond:2024ozy}
	P.~Ramond,
	\newblock \emph{{On the integrability of extended test body dynamics around black holes}},
	\newblock Class. Quant. Grav. \textbf{42}(6), 065019 (2025),
	\newblock \doi{10.1088/1361-6382/adb197},
	\newblock \eprint{2402.02670}.
	
	\bibitem{Harte:2026ihj}
	A.~I. Harte and P.~Ramond,
	\newblock \emph{{Octupole moments and the non-universality of free-fall in general relativity}}  (2026),
	\newblock \eprint{2607.21314}.
	
	\bibitem{Mogull:2020sak}
	G.~Mogull, J.~Plefka and J.~Steinhoff,
	\newblock \emph{{Classical black hole scattering from a worldline quantum field theory}},
	\newblock JHEP \textbf{02}, 048 (2021),
	\newblock \doi{10.1007/JHEP02(2021)048},
	\newblock \eprint{2010.02865}.
	
	\bibitem{Jakobsen:2023ndj}
	G.~U. Jakobsen, G.~Mogull, J.~Plefka, B.~Sauer and Y.~Xu,
	\newblock \emph{{Conservative Scattering of Spinning Black Holes at Fourth Post-Minkowskian Order}},
	\newblock Phys. Rev. Lett. \textbf{131}(15), 151401 (2023),
	\newblock \doi{10.1103/PhysRevLett.131.151401},
	\newblock \eprint{2306.01714}.
	
	\bibitem{Jakobsen:2023pvx}
	G.~U. Jakobsen, G.~Mogull, J.~Plefka and B.~Sauer,
	\newblock \emph{{Tidal effects and renormalization at fourth post-Minkowskian order}},
	\newblock Phys. Rev. D \textbf{109}(4), L041504 (2024),
	\newblock \doi{10.1103/PhysRevD.109.L041504},
	\newblock \eprint{2312.00719}.
	
	\bibitem{Jakobsen:2022psy}
	G.~U. Jakobsen, G.~Mogull, J.~Plefka and B.~Sauer,
	\newblock \emph{{All things retarded: radiation-reaction in worldline quantum field theory}},
	\newblock JHEP \textbf{10}, 128 (2022),
	\newblock \doi{10.1007/JHEP10(2022)128},
	\newblock \eprint{2207.00569}.
	
	\bibitem{Jakobsen:2023hig}
	G.~U. Jakobsen, G.~Mogull, J.~Plefka and B.~Sauer,
	\newblock \emph{{Dissipative Scattering of Spinning Black Holes at Fourth Post-Minkowskian Order}},
	\newblock Phys. Rev. Lett. \textbf{131}(24), 241402 (2023),
	\newblock \doi{10.1103/PhysRevLett.131.241402},
	\newblock \eprint{2308.11514}.
	
	\bibitem{Driesse:2024xad}
	M.~Driesse, G.~U. Jakobsen, G.~Mogull, J.~Plefka, B.~Sauer and J.~Usovitsch,
	\newblock \emph{{Conservative Black Hole Scattering at Fifth Post-Minkowskian and First Self-Force Order}},
	\newblock Phys. Rev. Lett. \textbf{132}(24), 241402 (2024),
	\newblock \doi{10.1103/PhysRevLett.132.241402},
	\newblock \eprint{2403.07781}.
	
	\bibitem{Driesse:2024feo}
	M.~Driesse, G.~U. Jakobsen, A.~Klemm, G.~Mogull, C.~Nega, J.~Plefka, B.~Sauer and J.~Usovitsch,
	\newblock \emph{{Emergence of Calabi{\textendash}Yau manifolds in high-precision black-hole scattering}},
	\newblock Nature \textbf{641}(8063), 603 (2025),
	\newblock \doi{10.1038/s41586-025-08984-2},
	\newblock \eprint{2411.11846}.
	
	\bibitem{Jakobsen:2021smu}
	G.~U. Jakobsen, G.~Mogull, J.~Plefka and J.~Steinhoff,
	\newblock \emph{{Classical Gravitational Bremsstrahlung from a Worldline Quantum Field Theory}},
	\newblock Phys. Rev. Lett. \textbf{126}(20), 201103 (2021),
	\newblock \doi{10.1103/PhysRevLett.126.201103},
	\newblock \eprint{2101.12688}.
	
	\bibitem{Jakobsen:2021zvh}
	G.~U. Jakobsen, G.~Mogull, J.~Plefka and J.~Steinhoff,
	\newblock \emph{{SUSY in the sky with gravitons}},
	\newblock JHEP \textbf{01}, 027 (2022),
	\newblock \doi{10.1007/JHEP01(2022)027},
	\newblock \eprint{2109.04465}.
	
	\bibitem{Driesse:2026qiz}
	M.~Driesse, G.~U. Jakobsen, G.~Mogull, C.~Nega, J.~Plefka, B.~Sauer and J.~Usovitsch,
	\newblock \emph{{Conservative Black Hole Scattering at Fifth Post-Minkowskian and Second Self-Force Order}},
	\newblock Phys. Rev. Lett. \textbf{137}(8), 081402 (2026),
	\newblock \doi{10.1103/4dvk-nglx},
	\newblock \eprint{2601.16256}.
	
	\bibitem{Haddad:2025cmw}
	K.~Haddad, G.~U. Jakobsen, G.~Mogull and J.~Plefka,
	\newblock \emph{{Unitarity and the On-Shell Action of Worldline Quantum Field Theory}}  (2025),
	\newblock \eprint{2510.00988}.
	
	\bibitem{Mogull:2025cfn}
	G.~Mogull, J.~Plefka and K.~Stoldt,
	\newblock \emph{{Radiated angular momentum from spinning black hole scattering trajectories}},
	\newblock Phys. Rev. D \textbf{112}(12), 124076 (2025),
	\newblock \doi{10.1103/my38-14k5},
	\newblock \eprint{2506.20643}.
	
	\bibitem{Cheung:2018wkq}
	C.~Cheung, I.~Z. Rothstein and M.~P. Solon,
	\newblock \emph{{From Scattering Amplitudes to Classical Potentials in the Post-Minkowskian Expansion}},
	\newblock Phys. Rev. Lett. \textbf{121}(25), 251101 (2018),
	\newblock \doi{10.1103/PhysRevLett.121.251101},
	\newblock \eprint{1808.02489}.
	
	\bibitem{Bern:2019nnu}
	Z.~Bern, C.~Cheung, R.~Roiban, C.-H. Shen, M.~P. Solon and M.~Zeng,
	\newblock \emph{{Scattering Amplitudes and the Conservative Hamiltonian for Binary Systems at Third Post-Minkowskian Order}},
	\newblock Phys. Rev. Lett. \textbf{122}(20), 201603 (2019),
	\newblock \doi{10.1103/PhysRevLett.122.201603},
	\newblock \eprint{1901.04424}.
	
	\bibitem{Bern:2022kto}
	Z.~Bern, D.~Kosmopoulos, A.~Luna, R.~Roiban and F.~Teng,
	\newblock \emph{{Binary Dynamics through the Fifth Power of Spin at O(G2)}},
	\newblock Phys. Rev. Lett. \textbf{130}(20), 201402 (2023),
	\newblock \doi{10.1103/PhysRevLett.130.201402},
	\newblock \eprint{2203.06202}.
	
	\bibitem{Bern:2024adl}
	Z.~Bern, E.~Herrmann, R.~Roiban, M.~S. Ruf, A.~V. Smirnov, V.~A. Smirnov and M.~Zeng,
	\newblock \emph{{Amplitudes, supersymmetric black hole scattering at $ \mathcal{O}\left({G}^5\right) $, and loop integration}},
	\newblock JHEP \textbf{10}, 023 (2024),
	\newblock \doi{10.1007/JHEP10(2024)023},
	\newblock \eprint{2406.01554}.
	
	\bibitem{Kosower:2018adc}
	D.~A. Kosower, B.~Maybee and D.~O'Connell,
	\newblock \emph{{Amplitudes, Observables, and Classical Scattering}},
	\newblock JHEP \textbf{02}, 137 (2019),
	\newblock \doi{10.1007/JHEP02(2019)137},
	\newblock \eprint{1811.10950}.
	
	\bibitem{Bjerrum-Bohr:2018xdl}
	N.~E.~J. Bjerrum-Bohr, P.~H. Damgaard, G.~Festuccia, L.~Plant{\'e} and P.~Vanhove,
	\newblock \emph{{General Relativity from Scattering Amplitudes}},
	\newblock Phys. Rev. Lett. \textbf{121}(17), 171601 (2018),
	\newblock \doi{10.1103/PhysRevLett.121.171601},
	\newblock \eprint{1806.04920}.
	
	\bibitem{Cristofoli:2019neg}
	A.~Cristofoli, N.~E.~J. Bjerrum-Bohr, P.~H. Damgaard and P.~Vanhove,
	\newblock \emph{{Post-Minkowskian Hamiltonians in general relativity}},
	\newblock Phys. Rev. D \textbf{100}(8), 084040 (2019),
	\newblock \doi{10.1103/PhysRevD.100.084040},
	\newblock \eprint{1906.01579}.
	
	\bibitem{Damgaard:2019lfh}
	P.~H. Damgaard, K.~Haddad and A.~Helset,
	\newblock \emph{{Heavy Black Hole Effective Theory}},
	\newblock JHEP \textbf{11}, 070 (2019),
	\newblock \doi{10.1007/JHEP11(2019)070},
	\newblock \eprint{1908.10308}.
	
	\bibitem{Brandhuber:2021eyq}
	A.~Brandhuber, G.~Chen, G.~Travaglini and C.~Wen,
	\newblock \emph{{Classical gravitational scattering from a gauge-invariant double copy}},
	\newblock JHEP \textbf{10}, 118 (2021),
	\newblock \doi{10.1007/JHEP10(2021)118},
	\newblock \eprint{2108.04216}.
	
	\bibitem{Bern:2021dqo}
	Z.~Bern, J.~Parra-Martinez, R.~Roiban, M.~S. Ruf, C.-H. Shen, M.~P. Solon and M.~Zeng,
	\newblock \emph{{Scattering Amplitudes and Conservative Binary Dynamics at ${\cal O}(G^4)$}},
	\newblock Phys. Rev. Lett. \textbf{126}(17), 171601 (2021),
	\newblock \doi{10.1103/PhysRevLett.126.171601},
	\newblock \eprint{2101.07254}.
	
	\bibitem{Bern:2021yeh}
	Z.~Bern, J.~Parra-Martinez, R.~Roiban, M.~S. Ruf, C.-H. Shen, M.~P. Solon and M.~Zeng,
	\newblock \emph{{Scattering Amplitudes, the Tail Effect, and Conservative Binary Dynamics at O(G4)}},
	\newblock Phys. Rev. Lett. \textbf{128}(16), 161103 (2022),
	\newblock \doi{10.1103/PhysRevLett.128.161103},
	\newblock \eprint{2112.10750}.
	
	\bibitem{Damgaard:2023ttc}
	P.~H. Damgaard, E.~R. Hansen, L.~Plant{\'e} and P.~Vanhove,
	\newblock \emph{{Classical observables from the exponential representation of the gravitational S-matrix}},
	\newblock JHEP \textbf{09}, 183 (2023),
	\newblock \doi{10.1007/JHEP09(2023)183},
	\newblock \eprint{2307.04746}.
	
	\bibitem{Bern:2023ccb}
	Z.~Bern, E.~Herrmann, R.~Roiban, M.~S. Ruf, A.~V. Smirnov, V.~A. Smirnov and M.~Zeng,
	\newblock \emph{{Conservative binary dynamics at order $O(\alpha^5)$ in electrodynamics}}  (2023),
	\newblock \eprint{2305.08981}.
	
	\bibitem{Bern:2025zno}
	Z.~Bern, E.~Herrmann, R.~Roiban, M.~S. Ruf, A.~V. Smirnov, V.~A. Smirnov and M.~Zeng,
	\newblock \emph{{Second-order self-force potential-region binary dynamics at $O(G^5)$ in supergravity}}  (2025),
	\newblock \eprint{2509.17412}.
	
	\bibitem{Herderschee:2023fxh}
	A.~Herderschee, R.~Roiban and F.~Teng,
	\newblock \emph{{The sub-leading scattering waveform from amplitudes}},
	\newblock JHEP \textbf{06}, 004 (2023),
	\newblock \doi{10.1007/JHEP06(2023)004},
	\newblock \eprint{2303.06112}.
	
	\bibitem{Brandhuber:2023hhy}
	A.~Brandhuber, G.~R. Brown, G.~Chen, S.~De~Angelis, J.~Gowdy and G.~Travaglini,
	\newblock \emph{{One-loop gravitational bremsstrahlung and waveforms from a heavy-mass effective field theory}},
	\newblock JHEP \textbf{06}, 048 (2023),
	\newblock \doi{10.1007/JHEP06(2023)048},
	\newblock \eprint{2303.06111}.
	
	\bibitem{Georgoudis:2023lgf}
	A.~Georgoudis, C.~Heissenberg and I.~Vazquez-Holm,
	\newblock \emph{{Inelastic exponentiation and classical gravitational scattering at one loop}},
	\newblock JHEP \textbf{2023}(06), 126 (2023),
	\newblock \doi{10.1007/JHEP06(2023)126},
	\newblock \eprint{2303.07006}.
	
	\bibitem{Cristofoli:2021vyo}
	A.~Cristofoli, R.~Gonzo, D.~A. Kosower and D.~O'Connell,
	\newblock \emph{{Waveforms from amplitudes}},
	\newblock Phys. Rev. D \textbf{106}(5), 056007 (2022),
	\newblock \doi{10.1103/PhysRevD.106.056007},
	\newblock \eprint{2107.10193}.
	
	\bibitem{Bohnenblust:2023qmy}
	L.~Bohnenblust, H.~Ita, M.~Kraus and J.~Schlenk,
	\newblock \emph{{Gravitational Bremsstrahlung in black-hole scattering at $ \mathcal{O}\left({G}^3\right) $: linear-in-spin effects}},
	\newblock JHEP \textbf{11}, 109 (2024),
	\newblock \doi{10.1007/JHEP11(2024)109},
	\newblock \eprint{2312.14859}.
	
	\bibitem{Bohnenblust:2025gir}
	L.~Bohnenblust, H.~Ita, M.~Kraus and J.~Schlenk,
	\newblock \emph{{Gravitational Bremsstrahlung in Black-Hole Scattering at $\mathcal{O}(G^3)$: Quadratic-in-Spin Effects}}  (2025),
	\newblock \eprint{2505.15724}.
	
	\bibitem{Parra-Martinez:2020dzs}
	J.~Parra-Martinez, M.~S. Ruf and M.~Zeng,
	\newblock \emph{{Extremal black hole scattering at $\mathcal{O}(G^3)$: graviton dominance, eikonal exponentiation, and differential equations}},
	\newblock JHEP \textbf{11}, 023 (2020),
	\newblock \doi{10.1007/JHEP11(2020)023},
	\newblock \eprint{2005.04236}.
	
	\bibitem{Weinzierl:2022eaz}
	S.~Weinzierl,
	\newblock \emph{{Feynman Integrals. A Comprehensive Treatment for Students and Researchers}},
	\newblock UNITEXT for Physics. Springer,
	\newblock ISBN 978-3-030-99557-7, 978-3-030-99560-7, 978-3-030-99558-4,
	\newblock \doi{10.1007/978-3-030-99558-4} (2022), \eprint{2201.03593}.
	
	\bibitem{Bern:1994cg}
	Z.~Bern, L.~J. Dixon, D.~C. Dunbar and D.~A. Kosower,
	\newblock \emph{{Fusing gauge theory tree amplitudes into loop amplitudes}},
	\newblock Nucl. Phys. B \textbf{435}, 59 (1995),
	\newblock \doi{10.1016/0550-3213(94)00488-Z},
	\newblock \eprint{hep-ph/9409265}.
	
	\bibitem{Bern:2011qt}
	Z.~Bern and Y.-t. Huang,
	\newblock \emph{{Basics of Generalized Unitarity}},
	\newblock J. Phys. A \textbf{44}, 454003 (2011),
	\newblock \doi{10.1088/1751-8113/44/45/454003},
	\newblock \eprint{1103.1869}.
	
	\bibitem{Bern:2010ue}
	Z.~Bern, J.~J.~M. Carrasco and H.~Johansson,
	\newblock \emph{{Perturbative Quantum Gravity as a Double Copy of Gauge Theory}},
	\newblock Phys. Rev. Lett. \textbf{105}, 061602 (2010),
	\newblock \doi{10.1103/PhysRevLett.105.061602},
	\newblock \eprint{1004.0476}.
	
	\bibitem{Bern:2019prr}
	Z.~Bern, J.~J. Carrasco, M.~Chiodaroli, H.~Johansson and R.~Roiban,
	\newblock \emph{{The Duality Between Color and Kinematics and its Applications}}  (2019),
	\newblock \eprint{1909.01358}.
	
	\bibitem{Bern:2022wqg}
	Z.~Bern, J.~J. Carrasco, M.~Chiodaroli, H.~Johansson and R.~Roiban,
	\newblock \emph{{The SAGEX review on scattering amplitudes Chapter 2: An invitation to color-kinematics duality and the double copy}},
	\newblock J. Phys. A \textbf{55}(44), 443003 (2022),
	\newblock \doi{10.1088/1751-8121/ac93cf},
	\newblock \eprint{2203.13013}.
	
	\bibitem{Chung:2018kqs}
	M.-Z. Chung, Y.-T. Huang, J.-W. Kim and S.~Lee,
	\newblock \emph{{The simplest massive S-matrix: from minimal coupling to Black Holes}},
	\newblock JHEP \textbf{04}, 156 (2019),
	\newblock \doi{10.1007/JHEP04(2019)156},
	\newblock \eprint{1812.08752}.
	
	\bibitem{Arkani-Hamed:2019ymq}
	N.~Arkani-Hamed, Y.-t. Huang and D.~O'Connell,
	\newblock \emph{{Kerr black holes as elementary particles}},
	\newblock JHEP \textbf{01}, 046 (2020),
	\newblock \doi{10.1007/JHEP01(2020)046},
	\newblock \eprint{1906.10100}.
	
	\bibitem{Bern:2020buy}
	Z.~Bern, A.~Luna, R.~Roiban, C.-H. Shen and M.~Zeng,
	\newblock \emph{{Spinning black hole binary dynamics, scattering amplitudes, and effective field theory}},
	\newblock Phys. Rev. D \textbf{104}(6), 065014 (2021),
	\newblock \doi{10.1103/PhysRevD.104.065014},
	\newblock \eprint{2005.03071}.
	
	\bibitem{Chiodaroli:2021eug}
	M.~Chiodaroli, H.~Johansson and P.~Pichini,
	\newblock \emph{{Compton black-hole scattering for s {\ensuremath{\leq}} 5/2}},
	\newblock JHEP \textbf{02}, 156 (2022),
	\newblock \doi{10.1007/JHEP02(2022)156},
	\newblock \eprint{2107.14779}.
	
	\bibitem{Aoude:2023vdk}
	R.~Aoude, K.~Haddad and A.~Helset,
	\newblock \emph{{Classical gravitational scattering at $\mathcal{O}(G^{2} S_{1}^{\infty} S_{2}^{\infty})$}},
	\newblock Phys. Rev. D \textbf{108}(2), 024050 (2023),
	\newblock \doi{10.1103/PhysRevD.108.024050},
	\newblock \eprint{2304.13740}.
	
	\bibitem{Cangemi:2023ysz}
	L.~Cangemi, M.~Chiodaroli, H.~Johansson, A.~Ochirov, P.~Pichini and E.~Skvortsov,
	\newblock \emph{{From higher-spin gauge interactions to Compton amplitudes for root-Kerr}},
	\newblock JHEP \textbf{09}, 196 (2024),
	\newblock \doi{10.1007/JHEP09(2024)196},
	\newblock \eprint{2311.14668}.
	
	\bibitem{Akpinar:2025bkt}
	D.~Akpinar, F.~Febres~Cordero, M.~Kraus, A.~Smirnov and M.~Zeng,
	\newblock \emph{{First Look at Quartic-in-Spin Binary Dynamics at Third Post-Minkowskian Order}},
	\newblock Phys. Rev. Lett. \textbf{135}(4), 041602 (2025),
	\newblock \doi{10.1103/c2dh-tj4v},
	\newblock \eprint{2502.08961}.
	
	\bibitem{FebresCordero:2022jts}
	F.~Febres~Cordero, M.~Kraus, G.~Lin, M.~S. Ruf and M.~Zeng,
	\newblock \emph{{Conservative Binary Dynamics with a Spinning Black Hole at O(G3) from Scattering Amplitudes}},
	\newblock Phys. Rev. Lett. \textbf{130}(2), 021601 (2023),
	\newblock \doi{10.1103/PhysRevLett.130.021601},
	\newblock \eprint{2205.07357}.
	
	\bibitem{Guevara:2019fsj}
	A.~Guevara, A.~Ochirov and J.~Vines,
	\newblock \emph{{Black-hole scattering with general spin directions from minimal-coupling amplitudes}},
	\newblock Phys. Rev. D \textbf{100}(10), 104024 (2019),
	\newblock \doi{10.1103/PhysRevD.100.104024},
	\newblock \eprint{1906.10071}.
	
	\bibitem{Guevara:2018wpp}
	A.~Guevara, A.~Ochirov and J.~Vines,
	\newblock \emph{{Scattering of Spinning Black Holes from Exponentiated Soft Factors}},
	\newblock JHEP \textbf{09}, 056 (2019),
	\newblock \doi{10.1007/JHEP09(2019)056},
	\newblock \eprint{1812.06895}.
	
	\bibitem{Kim:2021rda}
	J.-H. Kim, J.-W. Kim and S.~Lee,
	\newblock \emph{{The relativistic spherical top as a massive twistor}},
	\newblock J. Phys. A \textbf{54}(33), 335203 (2021),
	\newblock \doi{10.1088/1751-8121/ac11be},
	\newblock \eprint{2102.07063}.
	
	\bibitem{Kim:2024grz}
	J.-H. Kim, J.-W. Kim and S.~Lee,
	\newblock \emph{{Massive twistor worldline in electromagnetic fields}},
	\newblock JHEP \textbf{08}, 080 (2024),
	\newblock \doi{10.1007/JHEP08(2024)080},
	\newblock \eprint{2405.17056}.
	
	\bibitem{Cho:2023kux}
	K.~Cho, K.~Kim and K.~Lee,
	\newblock \emph{{Binary black holes and quantum off-shell recursion}},
	\newblock JHEP \textbf{05}, 050 (2024),
	\newblock \doi{10.1007/JHEP05(2024)050},
	\newblock \eprint{2311.01284}.
	
	\bibitem{Damgaard:2024fqj}
	P.~H. Damgaard and K.~Lee,
	\newblock \emph{{Schwarzschild Black Hole from Perturbation Theory to All Orders}},
	\newblock Phys. Rev. Lett. \textbf{132}(25), 251603 (2024),
	\newblock \doi{10.1103/PhysRevLett.132.251603},
	\newblock \eprint{2403.13216}.
	
	\bibitem{Bern:2026oqp}
	Z.~Bern, A.~Jackman, G.~Mansfield and M.~S. Ruf,
	\newblock \emph{{Classical Gravitational Scattering from the Ultraviolet and the Absence of Calabi-Yau Integrals in the Conservative Sector at $O(G^5)$}}  (2026),
	\newblock \eprint{2603.15383}.
	
	\bibitem{Dixon:1970mpd}
	W.~G. Dixon,
	\newblock \emph{Dynamics of extended bodies in general relativity. i. momentum and angular momentum},
	\newblock Proc. R. Soc. Lond. A \textbf{314}, 499 (1970),
	\newblock \doi{10.1098/rspa.1970.0020}.
	
	\bibitem{dixon2}
	W.~G. Dixon,
	\newblock \emph{Dynamics of extended bodies in general relativity - ii. moments of the charge-current vector},
	\newblock Proceedings of the Royal Society of London. A. Mathematical and Physical Sciences \textbf{319}(1539), 509 (1970),
	\newblock \doi{10.1098/rspa.1970.0191}.
	
	\bibitem{dixon3}
	W.~G. Dixon,
	\newblock \emph{Dynamics of extended bodies in general relativity iii. equations of motion},
	\newblock Philosophical Transactions of the Royal Society of London. Series A, Mathematical and Physical Sciences \textbf{277}(1264), 59 (1974),
	\newblock \doi{10.1098/rsta.1974.0046}.
	
	\bibitem{mpdreview}
	T.~Scheopner and F.~Lichtner,
	\newblock \emph{{Variational Approach to the Mathisson-Papapetrou-Dixon Equations of Motion for Extended Objects in General Relativity}},
	\newblock Submitted simultaneously to this article .
	
	\bibitem{Israel:1970kp}
	W.~Israel,
	\newblock \emph{{Source of the kerr metric}},
	\newblock Phys. Rev. D \textbf{2}, 641 (1970),
	\newblock \doi{10.1103/PhysRevD.2.641}.
	
	\bibitem{Beiglbock:1965fun}
	W.~Beiglböck,
	\newblock \emph{The centre-of-mass hypothesis in einstein's theory of gravitation},
	\newblock In \emph{International Conference on Relativistic Theories of Gravitation} (1965).
	
	\bibitem{Costa:2014nta}
	L.~F.~O. Costa and J.~Nat\'ario,
	\newblock \emph{{Center of mass, spin supplementary conditions, and the momentum of spinning particles}},
	\newblock Fund. Theor. Phys. \textbf{179}, 215 (2015),
	\newblock \doi{10.1007/978-3-319-18335-0_6},
	\newblock \eprint{1410.6443}.
	
	\bibitem{JanSteinhoff:2015ist}
	J.~Steinhoff,
	\newblock \emph{Spin and Quadrupole Contributions to the Motion of Astrophysical Binaries}, pp. 615--649,
	\newblock Springer International Publishing, Cham,
	\newblock ISBN 978-3-319-18335-0,
	\newblock \doi{10.1007/978-3-319-18335-0_19} (2015), \eprint{1412.3251}.
	
	\bibitem{Saketh:2022wap}
	M.~V.~S. Saketh and J.~Vines,
	\newblock \emph{{Scattering of gravitational waves off spinning compact objects with an effective worldline theory}},
	\newblock Phys. Rev. D \textbf{106}(12), 124026 (2022),
	\newblock \doi{10.1103/PhysRevD.106.124026},
	\newblock \eprint{2208.03170}.
	
	\bibitem{Arkani-Hamed:2017jhn}
	N.~Arkani-Hamed, T.-C. Huang and Y.-t. Huang,
	\newblock \emph{{Scattering amplitudes for all masses and spins}},
	\newblock JHEP \textbf{11}, 070 (2021),
	\newblock \doi{10.1007/JHEP11(2021)070},
	\newblock \eprint{1709.04891}.
	
	\bibitem{Bjerrum-Bohr:2026low}
	N.~E.~J. Bjerrum-Bohr, G.~Chen, K.~Papadimos, N.~Shah and C.~Su,
	\newblock \emph{{The Effective Kerr-Compton Tree Amplitude from Black Hole Perturbation Theory}}  (2026),
	\newblock \eprint{2610.40101}.
	
	\bibitem{Bautista:2023sdf}
	Y.~F. Bautista, G.~Bonelli, C.~Iossa, A.~Tanzini and Z.~Zhou,
	\newblock \emph{{Black hole perturbation theory meets CFT2: Kerr-Compton amplitudes from Nekrasov-Shatashvili functions}},
	\newblock Phys. Rev. D \textbf{109}(8), 084071 (2024),
	\newblock \doi{10.1103/PhysRevD.109.084071},
	\newblock \eprint{2312.05965}.
	
	\bibitem{Bautista:2024agp}
	Y.~F. Bautista, M.~Khalil, M.~Sergola, C.~Kavanagh and J.~Vines,
	\newblock \emph{{Post-Newtonian observables for aligned-spin binaries to sixth order in spin from gravitational self-force and Compton amplitudes}},
	\newblock Phys. Rev. D \textbf{110}(12), 124005 (2024),
	\newblock \doi{10.1103/PhysRevD.110.124005},
	\newblock \eprint{2408.01871}.
	
	\bibitem{Aoude:2022trd}
	R.~Aoude, K.~Haddad and A.~Helset,
	\newblock \emph{{Searching for Kerr in the 2PM amplitude}},
	\newblock JHEP \textbf{07}, 072 (2022),
	\newblock \doi{10.1007/JHEP07(2022)072},
	\newblock \eprint{2203.06197}.
	
	\bibitem{Cangemi:2023bpe}
	L.~Cangemi, M.~Chiodaroli, H.~Johansson, A.~Ochirov, P.~Pichini and E.~Skvortsov,
	\newblock \emph{{Compton Amplitude for Rotating Black Hole from QFT}},
	\newblock Phys. Rev. Lett. \textbf{133}(7), 071601 (2024),
	\newblock \doi{10.1103/PhysRevLett.133.071601},
	\newblock \eprint{2312.14913}.
	
	\bibitem{Bjerrum-Bohr:2023iey}
	N.~E.~J. Bjerrum-Bohr, G.~Chen and M.~Skowronek,
	\newblock \emph{{Covariant Compton Amplitudes in Gravity with Classical Spin}},
	\newblock Phys. Rev. Lett. \textbf{132}(19), 191603 (2024),
	\newblock \doi{10.1103/PhysRevLett.132.191603},
	\newblock \eprint{2309.11249}.
	
	\bibitem{Akpinar:2025tct}
	D.~Akpinar, G.~R. Brown, R.~Gonzo and M.~Zeng,
	\newblock \emph{{Unexpected Symmetries of Kerr Black Hole Scattering}}  (2025),
	\newblock \eprint{2508.10761}.
	
	\bibitem{Akpinar:2024meg}
	D.~Akpinar, F.~Febres~Cordero, M.~Kraus, M.~S. Ruf and M.~Zeng,
	\newblock \emph{{Spinning Black Hole Scattering at $\mathcal{O}(G^3 S^2)$: Casimir Terms, Radial Action and Hidden Symmetry}}  (2024),
	\newblock \eprint{2407.19005}.
	
	\bibitem{Aoude:2025xxq}
	R.~Aoude and A.~Helset,
	\newblock \emph{{Hidden simplicity in the scattering for neutron stars and black holes}}  (2025),
	\newblock \eprint{2509.04425}.
	
	\bibitem{Guevara:2020xjx}
	A.~Guevara, B.~Maybee, A.~Ochirov, D.~O'connell and J.~Vines,
	\newblock \emph{{A worldsheet for Kerr}},
	\newblock JHEP \textbf{03}, 201 (2021),
	\newblock \doi{10.1007/JHEP03(2021)201},
	\newblock \eprint{2012.11570}.
	
	\bibitem{Ben-Shahar:2025tiz}
	M.~Ben-Shahar, L.~Cangemi and H.~Johansson,
	\newblock \emph{{Kerr worldline-QFT action from Compton amplitude to infinite spin orders}}  (2025),
	\newblock \eprint{2512.24549}.
	
	\bibitem{Alessio:2025nzd}
	F.~Alessio, P.~Di~Vecchia, M.~Firrotta and P.~Pichini,
	\newblock \emph{{Searching for Kerr in string amplitudes}}  (2025),
	\newblock \eprint{2506.15529}.
	
	\bibitem{Bohnenblust:2024hkw}
	L.~Bohnenblust, L.~Cangemi, H.~Johansson and P.~Pichini,
	\newblock \emph{{Binary Kerr black-hole scattering at 2PM from quantum higher-spin Compton}}  (2024),
	\newblock \eprint{2410.23271}.
	
	\bibitem{Cangemi:2022bew}
	L.~Cangemi, M.~Chiodaroli, H.~Johansson, A.~Ochirov, P.~Pichini and E.~Skvortsov,
	\newblock \emph{{Kerr Black Holes From Massive Higher-Spin Gauge Symmetry}},
	\newblock Phys. Rev. Lett. \textbf{131}(22), 221401 (2023),
	\newblock \doi{10.1103/PhysRevLett.131.221401},
	\newblock \eprint{2212.06120}.
	
	\bibitem{1983mtbh.book.....C}
	S.~{Chandrasekhar},
	\newblock \emph{{The mathematical theory of black holes}} (1983).
	
	\bibitem{vanVoorthuizen:2024goj}
	J.~van Voorthuizen,
	\newblock \emph{{Multipole moments in stationary spacetimes}},
	\newblock J. Math. Phys. \textbf{66}(7), 072502 (2025),
	\newblock \doi{10.1063/5.0256660},
	\newblock \eprint{2412.19667}.
	
\end{thebibliography}
\end{document}